\documentclass[twocolumn]{aastex631} 
\usepackage{CJK}
\usepackage{graphicx,subfigure,amsmath,amsfonts,amssymb,multirow,mathrsfs,ulem,threeparttable}
\usepackage{pifont}
\usepackage{longtable}
\usepackage{url}
\usepackage{multirow} 

\begin{document}
\begin{CJK*}{UTF8}{gbsn}

\title{Can Dark Stars account for the star formation efficiency excess at very high redshifts?}

\author[0000-0003-4631-1915]{Lei Lei (雷磊)}
\affiliation{Key Laboratory of Dark Matter and Space Astronomy, Purple Mountain Observatory, Chinese Academy of Sciences, Nanjing 210023, China}
\affiliation{School of Astronomy and Space Science, University of Science and Technology of China, Hefei 230026, China}

\author[0000-0003-1215-6443]{Yi-Ying Wang (王艺颖)}
\affiliation{Key Laboratory of Dark Matter and Space Astronomy, Purple Mountain Observatory, Chinese Academy of Sciences, Nanjing 210023, China}
\affiliation{School of Astronomy and Space Science, University of Science and Technology of China, Hefei 230026, China}

\author[0000-0002-4538-8526]{Guan-Wen Yuan (袁官文)}
\affiliation{Department of Physics, University of Trento, Via Sommarive 14, 38123 Povo (TN), Italy}
\affiliation{Trento Institute for Fundamental Physics and Applications (TIFPA)-INFN, Via Sommarive 14, 38123 Povo (TN), Italy}

\author[0009-0004-0029-6080]{Tong-Lin Wang (王彤琳)}
\affiliation{Key Laboratory of Dark Matter and Space Astronomy, Purple Mountain Observatory, Chinese Academy of Sciences, Nanjing 210023, China}

\author[0000-0003-2723-6075]{Martin A. T. Groenewegen}
\affiliation{Royal Observatory of Belgium, Ringlaan 3, 1180 Brussels, Belgium}

\author[0000-0002-8966-6911]{Yi-Zhong Fan (范一中)}
\affiliation{Key Laboratory of Dark Matter and Space Astronomy, Purple Mountain Observatory, Chinese Academy of Sciences, Nanjing 210023, China}
\affiliation{School of Astronomy and Space Science, University of Science and Technology of China, Hefei 230026, China}
\correspondingauthor{Yi-Zhong Fan, Yi-Ying Wang}
\email{yzfan@pmo.ac.cn, wangyy@pmo.ac.cn}



\begin{abstract}

The James Webb Space Telescope (JWST) has recently conducted observations of massive galaxies at high redshifts, revealing a notable anomaly in their star formation efficiency (SFE). Motivated by the recent identification of three $\sim 10^{6}M_\odot$ dark star candidates, 
we investigate whether dark stars can be the origin of the SFE excess.
It turns out that the excess can be reproduced by a group of dark stars with $M \gtrsim 10^{3}\, \rm M_{\odot}$, because of their domination in generating primary UV radiation in high-redshift galaxies. The genesis of these dark stars is attributed to the capture of Weakly Interacting Massive Particles (WIMPs) within a mass range of tens of GeV to a few TeV. 
However, if the top-heavy initial mass function of dark stars holds up to $\sim 10^{5}M_\odot$, the relic black holes stemming from their collapse would be too abundant to be consistent with the current observations of Massive Compact Halo Objects (MACHOs). 
We thus suggest that just a small fraction of SFE excess may be contributed by the very massive dark stars and the majority likely originated from other reasons such as the Population III stars in view of their rather similar UV radiation efficiencies.

\end{abstract}


\section{Introduction}

The observations by JWST in the high redshift Universe have posed significant challenges to the prevailing $\Lambda$CDM model of galaxy formation \citep{2023Natur.616..266L,2023NatAs...7..731B,2024Natur.635..311X,2022ApJ...939L..31H,2024RAA....24d5001W,Menci:2024rbq}. These observations have raised two key challenges for theoretical frameworks explaining galaxy formation: Firstly, the elevated stellar mass density is observed in JWST high redshift galaxies at $z>6$~\citep{2023Natur.616..266L,2023NatAs...7..731B,2024arXiv240519953C,2024Natur.635..311X}. Secondly, the abundance of bright galaxies at $z>10$ is higher than the prediction of the ultraviolet luminosity functions (UV LF) under the $\Lambda$CDM framework \citep{2023ApJ...946L..13F,2024ApJ...965..169A,2023MNRAS.523.1009B,2023MNRAS.523.1036B,2023MNRAS.518.6011D,2024ApJ...960...56H,2024MNRAS.527.5004M,2023ApJ...946L..35M,2022ApJ...940L..14N,2023ApJ...951L...1P,2023MNRAS.518.2511L}. Both challenges can be attributed to an observed excess in star formation efficiency (SFE). The SFE excesses may be caused by the high formation efficiency of the first stars (or Population  \uppercase\expandafter{\romannumeral3}) in the early galaxies \citep{2024MNRAS.527.5929Y,2022ApJ...938L..10I} or several alternative scenarios, such as feedback-free star formation activities \citep{2023MNRAS.523.3201D}, dust-free star formation activities, or the presence of a stellar top-heavy initial mass function (IMF) \citep{2023ApJ...946L..13F,2024ApJ...960...56H} in the early Universe. 
However, these alternative models struggle to reconcile the observed high SFE approaching $100\%$ at $z\geq10$ in JWST observations.

Dark stars, which inhabit the first dark matter halos or mini halos in the high-redshift universe, are fueled by heating from dark matter \citep{Iocco:2008rb,Yoon:2008km,2008AIPC..990...42F,2012ApJ...761..154S,2016RPPh...79f6902F,Wu:2022wzw,Iocco:2024rez,Zhang:2023kfj}. This heating may originate from the gravitational attraction of dark matter and the annihilation of dark matter particles, particularly weakly interacting massive particles (WIMPs). These particles can potentially be captured through elastic scattering with baryonic matter \citep{Iocco:2010fxs,2010ApJ...716.1397F} and might be detectable with telescopic observations \citep{2010ApJ...717..257Z, 2010MNRAS.407L..74Z} . 
In the early Universe, dark matter densities were enhanced by a factor of $(1+z)^3$ at redshift $z$, suggesting a condition in favor of the formation of dark stars \citep{2016RPPh...79f6902F}. However, before the launch of JWST, identifying dark stars was challenging due to the lack of very high-redshift observations. Moreover, the outer atmospheric properties of dark stars show similar characteristics compared with those of stars undergoing nuclear reactions, enhancing the difficulty of identification.
The 
dark star observational features might include supermassive dark stars \citep{2008MNRAS.390.1655I,2009ApJ...692..574N,2010MNRAS.407L..74Z,2010ApJ...717..257Z,2012MNRAS.422.2164I,2016RPPh...79f6902F}, the extragalactic infrared background light \citep{2012ApJ...745..166M}, the extragalactic gamma-ray background \citep{2011JCAP...01..018S,2011JCAP...04..020Y,2012PhRvD..85h3519S}, remnant black holes following the demise of dark stars \citep{Ilie:2023aqu}, the influence of dark stars on the universe's reionization process \citep{2009PhRvD..79d3510S,2011ApJ...742..129S,2022ApJ...935...11G,2024PhRvD.109j3026Q} and so on. 

Recently, three potential supermassive dark star candidates have been identified by JWST~\citep{2023PNAS..12005762I}, for which the power source is rooted in the annihilation of dark matter particles rather than nuclear fusion~\citep{2016RPPh...79f6902F}. Supposing supermassive dark stars exist indeed, 
a population of dark stars living within the very high-redshift galaxies could influence the UV LF of galaxies. If the light output of these dark stars can be comparable with normal stars, the formation and evolution of galaxies will be affected in the early universe.
The study by \cite{Iocco:2024rez} estimated the UV limiting magnitudes for a massive dark star, approximately $10^6\, \rm M_{\odot}$ in size, situated within a dark matter halo having a mass of $M_h=10^8\, \rm M_{\odot}$ with a temperature $T=3.14\times 10^4 \,\rm  K$, and an accretion rate of $\dot{M_h}=10^{-1}\, \rm M_{\odot}\, yr^{-1}$. Their research concluded that the luminosity of the dar star is insufficient to account for the galaxy's brightness observed in JWST UV LF studies at $z\sim13$.  
In this work, we further investigate the potential contribution of the dark stars to the overall SFE.

In Section \ref{JWST+Obs}, we model the UV LF of high-redshift galaxies observed by the JWST 
considering both dark star and normal star populations. In Section \ref{MACHO}, we analyze the abundance of MACHOs and examine the possibility that they originated from
the collapse of dark stars. We then compare the UV LF model
with the UV LF data obtained from JWST. Section \ref{discussion} presents the conclusions and discussions 
The cosmological parameters used in this work include $H_{0}=67.36\, \rm km\, Mpc^{-1}\, s^{-1}$, $\Omega_{m}=0.3135$ and $\Omega_{\Lambda}=0.6847$ \citep{2020A&A...641A...6P}. The absolute bolometric (AB) magnitude system \citep{1983ApJ...266..713O} is adopted in this work.

\section{Observational Data and Fitting SFE Model}
\label{JWST+Obs}

\subsection{Observational Data}

To avoid the problem of overlapping skymaps, we have carefully selected the high-redshift UV LF data ($9\ge z \ge4$) from various sources. These sources encompass data from the Hubble Space Telescope \citep{2021AJ....162...47B}, the Subaru/Hyper Suprime-Cam survey \& CFHT Large Area U-band Survey \citep{2022ApJS..259...20H}, the Spitzer/Infrared Array Camera \citep{2020MNRAS.493.2059B} and the JWST \citep{2024ApJ...965..169A, 2023MNRAS.518.6011D, 2023ApJ...946L..35M, 2023ApJ...951L...1P}. 
For higher redshifts ($z\ge11$), where the luminosity contribution of dark stars is considered in our model, we adopt the UV LF data from relevant JWST literature \citep{2022ApJ...929....1H, 2023MNRAS.518.6011D, 2023MNRAS.523.1009B, 2023MNRAS.523.1036B, 2024ApJ...965...98C, 2023ApJS..265....5H, 2024MNRAS.527.5004M, 2023ApJ...951L...1P, 2024ApJ...969L...2F}. To ensure the reliability and coherence of our dataset, we have excluded any candidates identified as low redshift objects \citep{2023ApJ...954L..48W} and removed samples that are already covered in the JWST deep fields.

\subsection{SFE model}
\label{sect:SFE}
The star formation rate (SFR) can be determined from the UV luminosity using the relation~\citep{2014ARA&A..52..415M}
\begin{equation}\label{SFR_LUV}
{\rm SFR_{UV}}({\rm M}_{\odot} {\, \rm yr^{-1}})={\mathcal K}_{\rm UV} \times L_{\rm UV}(\rm erg \, s^{-1}\, Hz^{-1}).
\end{equation}
Here, ${\mathcal K}_{\rm UV}$ represents the conversion factor, which depends on the stellar populations of galaxies~\citep{2014ARA&A..52..415M}. In the standard scenario, this conversion factor takes the value ${\mathcal K}_{\rm UV}= 1.15\times10^{-28} \rm \,  M_{\odot}\, \rm yr^{-1} /(erg\, s^{-1}\, Hz^{-1})$, assuming a Salpeter IMF within the mass range of $0.1$ to $100\,\rm M_{\odot}$ \citep{1955ApJ...121..161S}. For extremely metal-poor ($Z=0$) Pop \uppercase\expandafter{\romannumeral3} stars characterized by a Salpeter IMF spanning $50$ to $500 \,  M_{\odot}$, the reported conversion factor decreases to
$2.80\times10^{-29} \rm \, \rm M_{\odot}\, yr^{-1} /(erg\, s^{-1}\, Hz^{-1})$~\citep{2022ApJ...938L..10I}. 
In our analysis, we estimate the conversion factor using the mass-to-luminosity ratio of dark stars, assuming a power-law IMF with $\phi(m) \propto m^{-0.17}$ (see Appendix \ref{app:UVLF} for details). We find the conversion factor for dark stars aligns closely with that of Pop \uppercase\expandafter{\romannumeral3} stars. Consequently, we adopt the conversion factor ${\mathcal K}_{\rm UV}= 2.80\times10^{-29} \rm \, \rm M_{\odot}\, yr^{-1} /(erg\, s^{-1}\, Hz^{-1})$ to fit the UV LF data.

We employ an extensive model of galaxy evolution to analyze the UV LF and explore the evolution of the SFE. The SFR is derived from the accretion rate of baryons and the total SFE, expressed as
\begin{equation}\label{eq+SFE_tot}
{\rm SFR_{UV}} = f_{\rm tot}\times \dot{M}_{b},~~~~
f_{\rm tot} = f_{\rm S}+f_{\rm DS},
\end{equation}
where $f_{\rm S}$ ($f_{\rm DS}$) represents the SFE of stars (dark stars), $\dot{M_b}$ is the baryon accretion rate of the galaxy, which can be derived from
the accretion rate of the dark matter halo: 
\begin{equation}
    \dot{M_{\rm b}} = f_{\rm b} \times \dot{M_{\rm h}}, 
\end{equation}
\begin{equation}
    f_{\rm b}\equiv \frac{\Omega_{\rm b}}{\Omega_{\rm m}} =0.156. 
\end{equation}
The accretion rate of the dark matter halo $\dot{M_{\rm h}}$ can be obtained by the simulation~\citep{2015ApJ...799...32B}. 

Due to the differences in physical processes, the formation efficiency of star $f_S$ and dark star $f_{DS}$ in Equation~(\ref{eq+SFE_tot}) may differ in definitions. Observations have indicated that stars form in the dense, cold, molecular phase of the ISM. Current detections support a (nearly) universal low star formation efficiency in Equation~(\ref{eq:SFE_S}) in the nearby galaxies 
and indicate that the temperature of dust/gas will influence SFE. The energy injection from radiation processes like reionization, stellar winds, supernovae and AGN will negatively affect the SFE $f_S$ \citep{2015ARA&A..53...51S,2018ARA&A..56..435W}. In this work, the star formation efficiency $f_S$ is calculated using a parametric formula dependent on the dark matter halo mass within the extensive model of galaxy evolution~\citep{2018ARA&A..56..435W}:
\begin{equation}\label{eq:SFE_S}
f_{\rm S}=\frac{2\epsilon_{\rm N} }{\bigl( \frac{M_{\rm h}}{M_1} \bigr )^{-\beta} + \bigl( \frac{M_{\rm h}}{M_1} \bigr)^{ \gamma}}.
\end{equation}
In this expression, $\epsilon_{\rm N}$ denotes the normalized constant, $M_1$ is the characteristic mass where the SFE is equal to $\epsilon_{\rm N}$, $M_{\rm h}$ is the halo mass, and $\beta, \gamma$ are slopes determining the decrease in SFE at low and high masses, respectively. The characteristic mass of the dark matter halo is set to be
$M_{\rm 1} = 10^{12}\, \rm M_{\odot}$ as suggested in ~\cite{2022ApJS..259...20H} and ~\cite{2023ApJ...954L..48W}. 

However, dark stars are thought to be powered by dark matter annihilation at the galaxy's center. Therefore, their formation efficiency is likely to be correlated with the mass of the dark matter halo strongly. 
Compared with normal stars, the formation efficiency of dark stars is more reliant on the density of dark matter, which is directly linked to the mass of the dark matter halo. Given the absence of mature simulations or analytical theories on this topic, it becomes necessary to define a new formation efficiency function specifically for dark stars. 
For characterizing the formation efficiency of dark stars, we utilize a power-law model expressed as:
\begin{equation}\label{eq:SFE_DS}
	f_{\rm DS}= \epsilon _{\rm DS} \bigl(
 \frac{M_{\rm h}}{M_1} \bigr)^{ \gamma _{\rm DS}},
\end{equation}
where $\epsilon_{\rm DS}$ is the normalization constant, $M_1$ is the characteristic mass, and $\gamma_{\rm DS}$ is the slope governing the dark star SFE across different halos masses. As shown in Equation~\ref{eq:SFE_DS}, the dark star formation efficiency $f_{DS}$ is assumed to have 
a monotonically incremental relationship with dark matter halo mass $M_h$. 

Pop III stars are believed to form in low metallicity gas clouds within dark matter minihalos \citep{2013ASSL..396..103G,2013ASSL..396..177J,2023ARA&A..61...65K}. However, the details remain to be better understood, and it is likely that the formation efficiency is related to the mass of the halo. Consequently, we assume that the formation efficiency of Pop III stars is analogous to that of dark stars, as outlined in Equation~(\ref{eq:SFE_DS}).
In our analysis, we utilize the conversion factor ${\mathcal K}_{\rm UV}= 2.80\times10^{-29} \rm \, \rm M_{\odot}\, yr^{-1} /(erg\, s^{-1}\, Hz^{-1})$ for both dark star and Pop III star formation efficiency models to fit the UV LF data. The primary distinction between the dark stars and Pop III stars in this study lies in their IMFs: Pop III stars have a mass range of $50\leq M_{*} \leq 500 \, \rm M_{\odot}$ with a Salpeter IMF $\phi(m)\propto m^{-2.35}$, while dark stars have a mass range exceeding $500\, \rm M_{\odot}$ and follow a top-heavy IMF $\phi(m)\propto m^{+0.17}$.

According to~\cite{2023ApJ...954L..48W}, the theoretical model of UV LFs can be written as
\begin{equation}\label{eq:UVLF}
	\Phi(M_{\rm UV})=\phi(M_{\rm h}) \, \bigg \vert \frac{{\rm d} M_{\rm h}}{{\rm d} M_{\rm UV}} \bigg \vert,
\end{equation}
where $\big \vert \frac{{\rm d} M_{\rm h}}{{\rm d} M_{\rm UV}} \big \vert$ is the Jacobi matrix mapping from $\phi(M_{\rm h})$ to $\Phi(M_{\rm UV})$, and $M_{\rm UV}$ is the dust-corrected or intrinsic magnitude, depending on whether the dust-attenuation effect is considered or not. The halo mass number  density function $\phi(M_{\rm h})$ is from the number of DM halos per unit mass per unit comoving volume:
\begin{equation}\label{eq:halo_model}
	\frac{{\rm d} n}{d \ln M_{\rm h}} = M_{\rm h} \cdot \frac{\rho_0}{M_{\rm h}^2} f(\sigma) \, \bigg \vert \frac{{\rm d} \ln \sigma}{{\rm d} \ln M_{\rm h}} \bigg \vert ,
\end{equation}
where $\rho_0$ is the mean density of the Universe and $\sigma$ is the r.m.s. variance of mass, which is determined by the linear power spectrum and the top-hat window function. The linear power spectrum can be computed using the transfer function provided by the public Code for Anisotropies in the Microwave Background (CAMB) \citep{2000ApJ...538..473L}. The mass function $f(\sigma)$ of DM halos was from the high-resolution N-body simulations in \cite{2007MNRAS.374....2R}. In detail, $\phi(M_{\rm h})$ can be derived by the public package {\sc HMFcalc} \citep{2013A&C.....3...23M} conveniently. Assuming $M_{\rm h}$ distributes in a wide range from $10^2$ to $10^{16} \, \rm M_{\odot}$ with a tiny bin ($\log_{10}{\rm \Delta} M_{\rm h} = 0.01$), the absolute magnitude ($M_{\rm UV}$) of the corresponding galaxy for each DM halo ($M_{\rm h}$) can be derived from Equations~(\ref{SFR_LUV})-(\ref{eq:SFE_DS}). Therefore, the term of $\frac{{\rm d}M_{\rm h}}{{\rm d}{M_{\rm UV}}}$ in Equation~(\ref{eq:UVLF}) can be calculated by the differentials of $M_{\rm h}$ and $M_{\rm UV}$. After building such a connection between $M_{\rm h}$ and $M_{\rm UV}$, the UV LFs model can be constructed by Equation~(\ref{eq:UVLF}).

\subsection{Fitting Results}

All of the variable parameters in Eq.~(\ref{eq+SFE_tot},\ref{eq:SFE_S},\ref{eq:SFE_DS}) can be estimated by fitting the observations of the UV luminosity functions with Eq.~(\ref{eq:UVLF}). In Bayesian analysis, the likelihood function follows an asymmetric normal distribution because of the asymmetric uncertainties of these real data:
\begin{equation}\label{eq:likelihood}
	\mathcal{L}=\prod_i^N \operatorname{AN}\left(f\left(x_i\right)-y_i \mid c_i, d_i\right) ,
\end{equation}
where AN is the asymmetric normal distribution, $f(x_i)$ are the observed UV LFs at magnitudes $x_i$, $y_i$ are the model values, $c_i$ and $d_i$ represent the deviation and the skewness of the distribution can be obtained from the asymmetric errors of the observation (i.e. the Equation (14) of \cite{2013ApJ...778...66K}). 
We use the nested sampling method and adopt \emph{Pymultinest}\citep{2014A&A...564A.125B} as the sampler to calculate the posterior distributions of the parameters.
In the \emph{Pymultinest} framework, the comparison of the models is assessed through the computation of the Bayesian evidence, as detailed by \cite{2009MNRAS.398.1601F,2014A&A...564A.125B}. The Bayesian evidence, denoted as \(\mathcal{Z}\), is given by the following integral:
\begin{equation}\label{eq:BayesEvidence}
	\mathcal{Z} = \int \mathcal{L}(\boldsymbol{\Theta}) \pi(\boldsymbol{\Theta}) d^D \boldsymbol{\Theta},
\end{equation}
where \(\boldsymbol{\Theta}\) represents the set of parameters within the model, \(\pi(\boldsymbol{\Theta})\) is the prior probability distribution, and \(D\) signifies the dimensionality of the parameter space.

\begin{figure}[h]
\centering
\includegraphics[width=0.99\linewidth]{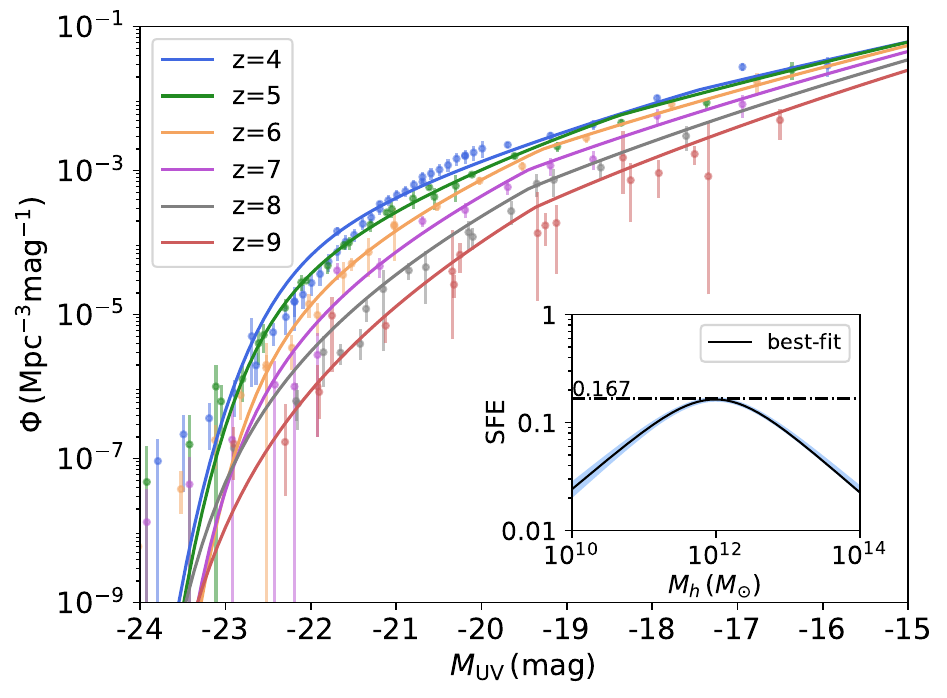}
\caption{\label{fig:UVLF_z4z9} 
The fitting results of UV LF observations within the redshift range of approximately $4-9$. All of the solid lines represent the optimal fit for the complete dataset, with data points derived from observations. In the subfigure, the light blue region illustrates the comprehensive posterior distribution of SFE.}  
\end{figure}

In the range of $4\leq z \leq 9$, we assume a constant $\epsilon_{\rm N}$ value for SFE to estimate the contribution of the Pop \uppercase\expandafter{\romannumeral2} star formation. When fitting the UV LFs, we only consider the Pop \uppercase\expandafter{\romannumeral2} star formation rate (i.e. discard the dark star part of Equation \ref{eq+SFE_tot}: ${\rm SFR_{UV}} = f_{\rm S}\times \dot{M}_{b}$, where the Pop \uppercase\expandafter{\romannumeral2} star formation efficiency $f_{\rm S}$ is defined by Equation \ref{eq:SFE_S}). As shown in Figure \ref{fig:UVLF_z4z9}, the SFE is constrained within a tight range, resulting in the well-fitted UV LFs. Table \ref{Tab:z4-9parameters} shows the best-fit values and posterior distributions of the parameters of the Pop \uppercase\expandafter{\romannumeral2} star formation efficiency model. The best-fit values of the parameters are taken from posteriors corresponding to the maximum likelihood. The $1\sigma$ range of the fitting error of the parameter in the tables is 68\% credible level of the posterior distribution. Subsequently, the best-fit values are used to calculate the SFE of Pop \uppercase\expandafter{\romannumeral2} stars at $z\ge10$.
The remaining contribution to the SFR is attributed to the formation of dark stars. In Eq.~(\ref{eq+SFE_tot}), we introduce $f_{\rm DS}$ to represent the efficiency of dark star formation. In the analysis, we specifically extrapolated the Pop \uppercase\expandafter{\romannumeral2} star formation contribution to the UV LF at redshifts $z \sim 11-14$ using the fitting results from the redshift range $z \sim 4-9$. The SFE parameters that describe the contribution of Pop \uppercase\expandafter{\romannumeral2} are fixed to the best-fit values at $z \sim 4-9$, which are listed in Table \ref{Tab:z4-9parameters}. 
The remaining contributions at $z \sim 11-14$ were then attributed to dark stars (or Pop \uppercase\expandafter{\romannumeral3}) and fitted using a power-law model specific to these populations. This method allowed us to isolate the contribution of dark stars (or Pop \uppercase\expandafter{\romannumeral3}) and assess their influence on the UV LF at these elevated redshifts.

\begin{table*}[hbtp]
\begin{ruledtabular}
\centering
\caption{The best-fit Values and Posterior Results of the Pop \uppercase\expandafter{\romannumeral2} SFE parameters at $4\leq z \leq 9$}
\label{Tab:z4-9parameters}
\begin{tabular}{c|ccc|ccc|c}
\multirow{2}{*}{redshift} &\multicolumn{3}{c|}{Best-fit Values} &\multicolumn{3}{c|}{Posterior Results at 68\% Credible Level} &\multirow{2}{*}{$\ln(\mathcal{Z})$} \\
&$\epsilon_{\rm N}$ &$\beta$ &$\gamma$  &$\epsilon_{\rm N}$ &$\beta$ &$\gamma$ \\ \hline     
4-9\textsuperscript{a} & 0.158 & 0.564 & 0.580 & $0.157^{+0.001}_{-0.001}$  & $0.564^{+0.008}_{-0.008}$ & $0.577^{+0.003}_{-0.004}$ & 1258.46 \\
\end{tabular}
\end{ruledtabular}
\begin{tablenotes}
 \item[a] \textsuperscript{a} In the range of $4\leq z \leq 9$, we fitted the data with a Pop \uppercase\expandafter{\romannumeral2} star formation efficiency without considering dark stars (or Pop \uppercase\expandafter{\romannumeral3} stars).
\end{tablenotes}
\end{table*}

\begin{figure*}[htbp]
\centering
\includegraphics[width=0.48\linewidth]{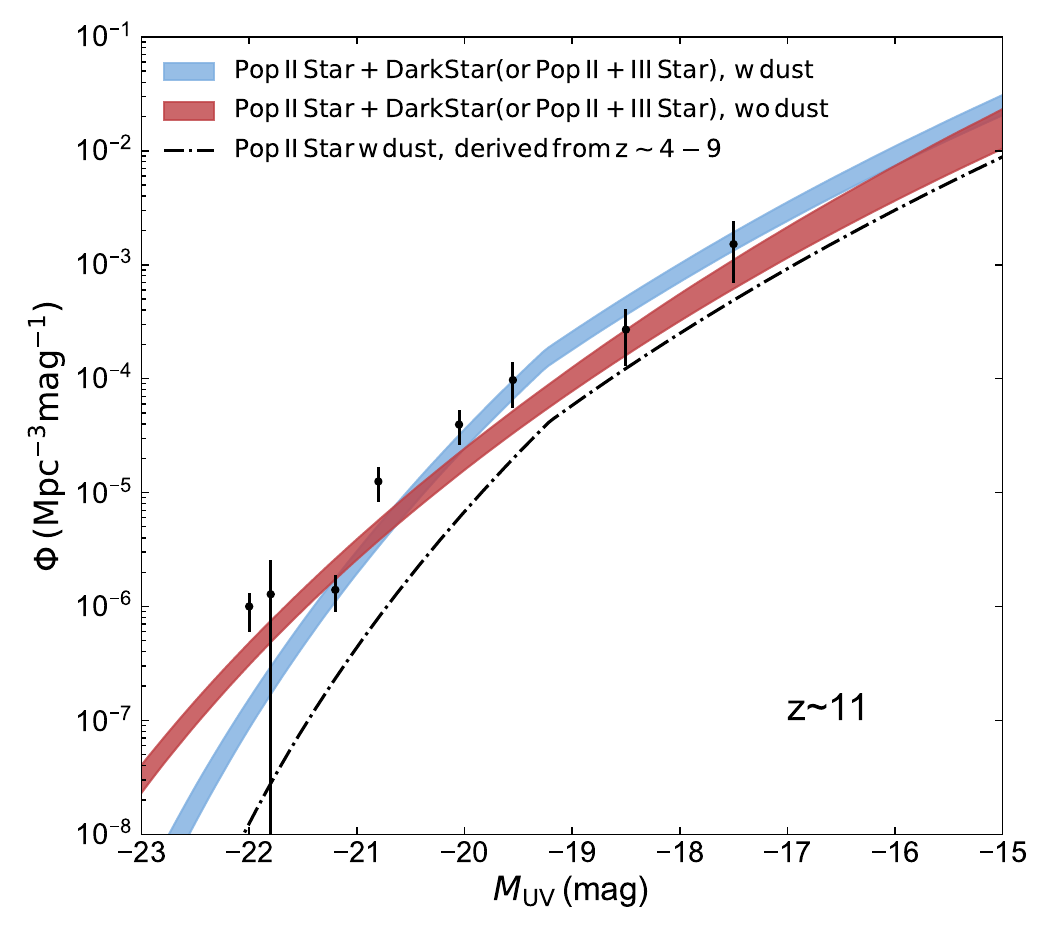}
\includegraphics[width=0.48\linewidth]{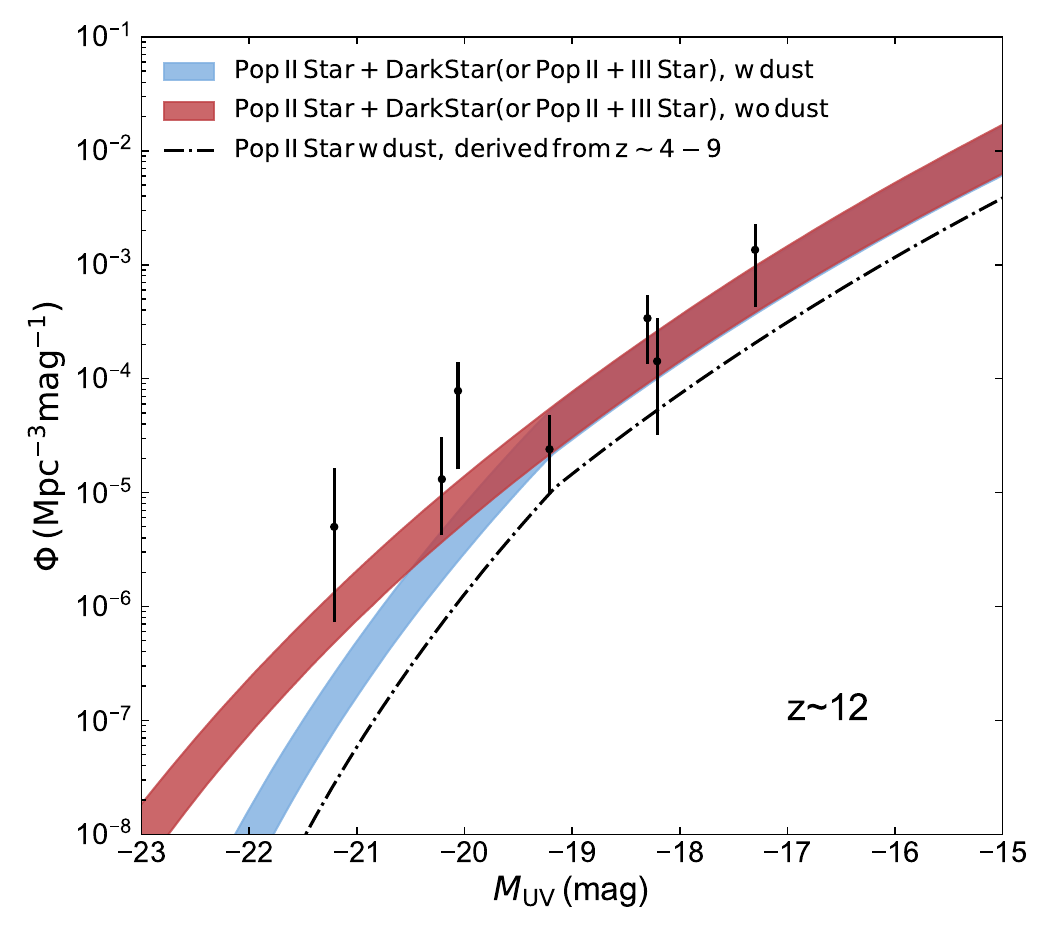}
\includegraphics[width=0.48\linewidth]{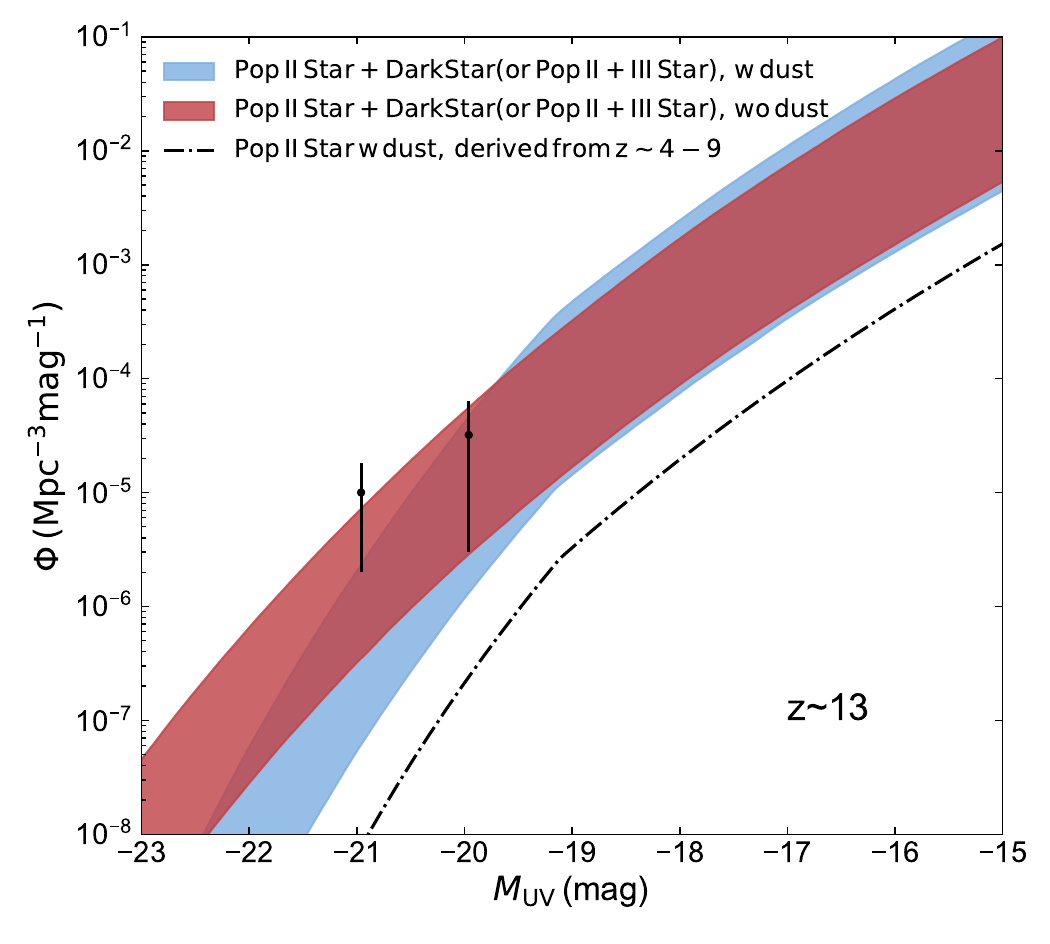}
\includegraphics[width=0.48\linewidth]{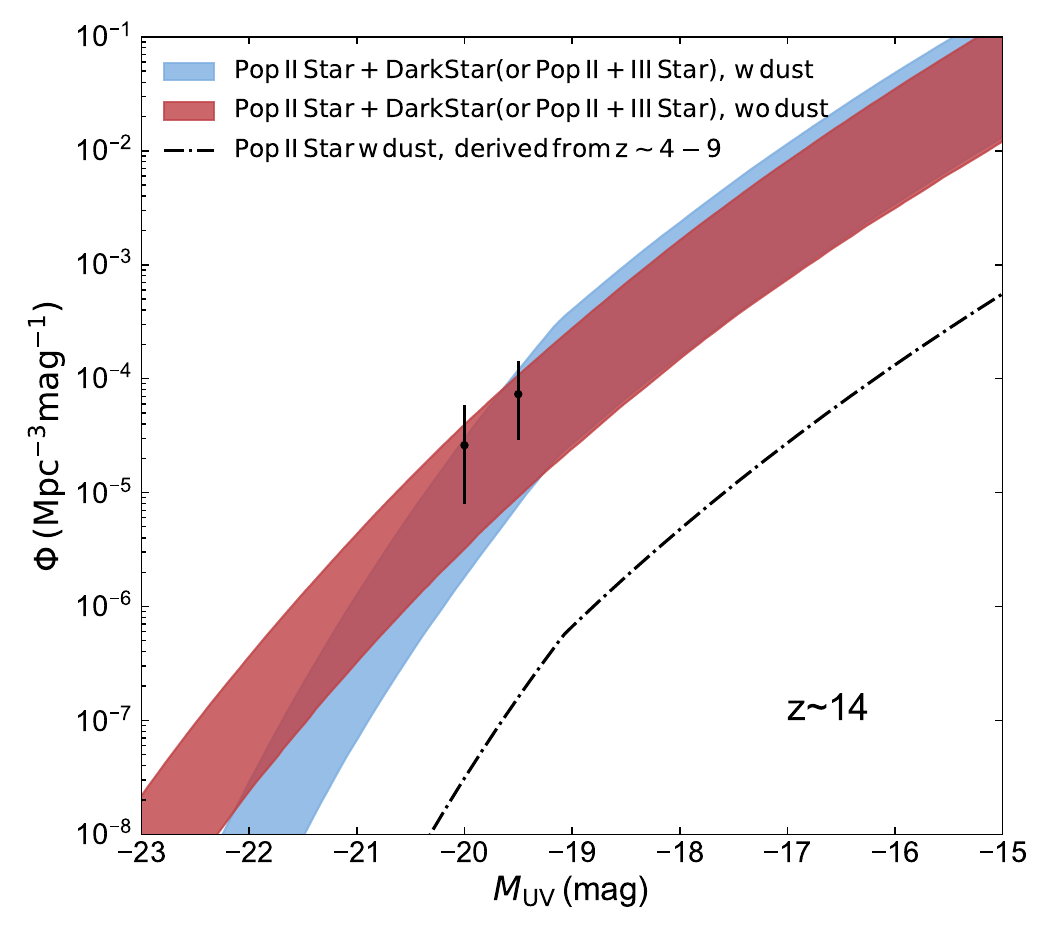}
\caption{\label{fig:UVLF} 
The UV LFs of galaxies at high redshift ($z>10$) as observed by JWST, alongside model predictions. The UV LFs data is compiled from diverse projects conducted with JWST~\citep{2022ApJ...929....1H, 2023MNRAS.518.6011D, 2023MNRAS.523.1009B, 2023MNRAS.523.1036B, 2024ApJ...965...98C, 2023ApJS..265....5H, 2024MNRAS.527.5004M, 2023ApJ...951L...1P}. The 68.3\% posterior regions of the Pop \uppercase\expandafter{\romannumeral2} Star+Dark Star (or Pop \uppercase\expandafter{\romannumeral2} + \uppercase\expandafter{\romannumeral3} Star) models, with and without dust, are represented by blue and red stripes, respectively. The black dash-dotted line represents the predicted high-redshift Pop \uppercase\expandafter{\romannumeral2} star model, which is derived from the extrapolation of low-redshift ($z\sim4-9$) fitting results. The star formation efficiency (SFE) models for both dark stars and Population III (Pop III) stars are assumed to be identical.   } 
\end{figure*}

Figure \ref{fig:UVLF} shows the UV LF and SFE models at redshifts of $z= 11, 12, 13, 14$. The blue and red bands depict the 68.3\% posterior regions of the Pop \uppercase\expandafter{\romannumeral2} Star+Dark Star (or Pop \uppercase\expandafter{\romannumeral2}+ \uppercase\expandafter{\romannumeral3} Star) models, including two cases with and without the effects of dust attenuation, respectively. We also applied the Pop \uppercase\expandafter{\romannumeral2} star+dark-star model to fit the UV LF at redshifts around \( z \sim 10 \). However, our results indicate that the SFE of the dark star (or Pop \uppercase\expandafter{\romannumeral3}) is a feeble contribution, which is $\sim 2$ dex lower than other redshifts ($z>10$). Because the $z\sim10$ UV LF can be fitted well without the dark star (or Pop \uppercase\expandafter{\romannumeral3} star), we show the UV LF data and fitted models at $z\sim 11-14$ in Figure~\ref{fig:UVLF}.  
The black dash-dotted line represents the UV LF model without dark stars (or Pop \uppercase\expandafter{\romannumeral3} stars), derived from the best-fitted UV LF model in the redshift bin $z\sim 4-9$.

The derived model lines are significantly lower than the high-redshift JWST UV LF data, indicating a significant transition from the low- to the high-redshift range. The tension between the UV LF model and the data becomes more pronounced at higher redshifts, reaching approximately $\sim2\, \rm dex$/$\sim3\, \rm dex$ at $z\sim13$/$z\sim14$, respectively. Specifically, the bright end of the UV LF exhibits a larger tension in each redshift bin. The Pop \uppercase\expandafter{\romannumeral2} Star+Dark Star (or Pop \uppercase\expandafter{\romannumeral2}+ \uppercase\expandafter{\romannumeral3} Star) scenario provides a good fit for the current data, with dark stars playing a dominant role. The higher UV radiative efficiency of dark stars leads to a lower SFE fitting the UV LF. The posterior distributions of the SFE model can be found in Figure~\ref{fig:SFE_withDS} of Appendix \ref{app:SFEpost}.

\begin{table}[hbtp]
\begin{ruledtabular}
\centering
\caption{The best-fit Values and Posterior Results of the Dark Star (or Pop \uppercase\expandafter{\romannumeral3} Star) SFE parameters at $11\leq z \leq 14$}
\label{Tab:z11-14parameters}
\begin{tabular}{c|c|c|c|c|c}
\multirow{2}{*}{z} &\multicolumn{2}{c|}{Best-fit} &\multicolumn{2}{c|}{Posterior\textsuperscript{a}} &\multirow{2}{*}{$\ln(\mathcal{Z})$}   \\
\cline{2-5}
& $\epsilon_{\rm DS}$ &$\gamma_{\rm DS}$ &  $\epsilon_{\rm DS}$ & $\gamma_{\rm DS}$   \\
\hline 
\multicolumn{6}{c}{with dust} \\
\hline 
11 &  0.076 & 0.520 & $0.059^{+0.013}_{-0.017}$ & $0.469^{+0.046}_{-0.075}$ &  79.76   \\
12 &  0.005 & 0.004  & $0.024^{+0.023}_{-0.017}$ &  $0.474^{+0.154}_{-0.188}$ &   58.75   \\
13 &  0.179 & 0.335 & $0.063^{+0.069}_{-0.045}$ & $0.293^{+0.196}_{-0.150}$ &  34.84  \\
14 &  0.201 & 0.296 & $0.103^{+0.066}_{-0.060}$ & $0.241^{+0.105}_{-0.120}$  &  18.49  \\
\hline 
\multicolumn{6}{c}{without dust} \\
\hline 
11 &  0.004 & 0.003 & $0.012^{+0.011}_{-0.007}$ &  $0.452^{+0.221}_{-0.282}$ &  79.40  \\
12 & 0.063 & 0.558 &  $0.025^{+0.024}_{-0.017}$ & $0.469^{+0.145}_{-0.187}$  &   59.29   \\
13 & 0.164 & 0.353 & $0.066^{+0.058}_{-0.045}$ & $0.309^{+0.162}_{-0.145}$  &  35.60  \\
14 & 0.080 & 0.161 & $0.090^{+0.064}_{-0.052}$ & $0.249^{+0.107}_{-0.129}$  &  18.52  \\
\end{tabular}
\end{ruledtabular}
\begin{tablenotes}
 \item[a] \textsuperscript{a} The listed errors of the posterior results are at 68\% Credible Level.
\end{tablenotes}
\end{table}

Table~\ref{Tab:z11-14parameters} presents the best-fit values of dark star (or Pop \uppercase\expandafter{\romannumeral3} star) formation parameter and the posterior results of the SFE model parameters. We also plot the corners of the posteriors in Figure \ref{fig:posterior} and \ref{fig:posteriorwodust} in the Appendix \ref{app:SFEpost}. The darkred regions of the three depths in the corner plot in Figure \ref{fig:posterior} and \ref{fig:posteriorwodust} represent 68\%, 95\%, and 99\% confidence intervals, respectively. The black crosses are the positions of the best-fit parameters. Even if the best-fit parameter is within the $68\%$ credible interval on the two-dimensional corner plot, the best-fit value of single parameter may deviated from $68\%$ credible interval because of the non-Gaussian posterior distribution.
The fitted dark star (or Pop \uppercase\expandafter{\romannumeral3} star) formation efficiencies, denoted as $\epsilon_{\rm DS}$, for the redshift bins around $z \sim 13$ and $z \sim 14$ are higher than those for the bins around $z \sim 11$ and $z \sim 12$. This trend is in agreement with the UV LF models depicted in Figure~\ref{fig:UVLF}. 
However, we also see that the number of UV LF data points for $z\sim 13$ and $z\sim 14$ is very small, so more data is needed for more robust investigations in the future.

The Bayesian evidence is a metric that quantifies the relative fit of two models to the data, with higher values indicating superior performance on the dataset. By examining the log-evidence values for the various models presented in Table \ref{Tab:z11-14parameters}, we are able to assess the impact of including or excluding dust attenuation in our UV LF model fits.
The Bayes factor ($\mathcal{B}$), defined as the ratio of the evidences for two models, is a standard tool for gauging the relative support for one model over another: $\mathcal{B} = \frac{\mathcal{Z}_a}{\mathcal{Z}_b}$.
As demonstrated, the differences in the log-evidences for our models, expressed as $\ln(\mathcal{B}) = \ln(\mathcal{Z}_a) - \ln(\mathcal{Z}_b)$, are less than 1.0 when comparing fits with and without consideration of dust. This suggests that our model fits are robust and not significantly influenced by the inclusion of dust extinction effects, as the Bayes factors do not exceed 3, indicating no strong evidence in favor of one model over the other.

\section{Constraint of MACHOs from Dark Stars' Collapse}
\label{MACHO}

According to the observations summarized in \cite{2021NatRP...3..732V}, there are two key facts regarding the low spatial density of massive black holes in the Universe: (1) the local Universe's abundance of massive black holes is estimated to be $n_{\rm BH} \sim 0.01 - 0.001 \rm Mpc^{-3}$, which is relatively low compared to the population of stars (also see \cite{2020ARA&A..58..257G}); (2) luminous quasars at redshifts around $z \sim 6$ are quite rare, with a density of $n_{\rm BH} \sim 10^{-9} \rm Mpc^{-3}$ (for the details please see \cite{2001AJ....122.2833F}). These observations suggest that massive black holes are not as prevalent as one might expect in the Universe's population of celestial objects.

In addition to the number density of massive black holes, other works constrain the proportion of black holes in the halo mass in galaxies. The constraints on the fraction of black holes in dark matter halo are more straightforward 
to limit dark star formation because the dark star formation efficiency in our work is described as a power-law model connected with the dark matter halo mass function. Thus we searched for possible constraints on the BH fraction in the halo. Three constraints are suitable to this work because of the mass range and physical properties: (1) constraint on black holes or compact dark matter with microlensing events near the cluster strong lensing critical curves \citep{2018PhRvD..97b3518O}; (2) constraint from MACHO dark matter will dynamically heat the star cluster near the center of the ultra-faint dwarf galaxy \citep{2016ApJ...824L..31B}; (3) constraints from heat transfer between MACHOs and stars caused by gravitational scattering  \citep{2023arXiv231107654G}.

The Pop III stars within the mass range of \(140 \leq M_{*} \leq 260 \, \rm M_{\odot}\) \citep{2024MNRAS.527.5102V} will not form black holes. Instead, they will experience the pair-instability supernova process, which leads to the ejection of most of their mass. However, dark stars with higher masses (\(\geq 500 \, \rm M_{\odot}\)) will contribute to the population of Massive Astrophysical Compact Halo Objects (MACHOs) and provide minimal mass feedback to their surroundings.

\begin{figure}[htbp]
\centering
\includegraphics[width=0.99\linewidth]{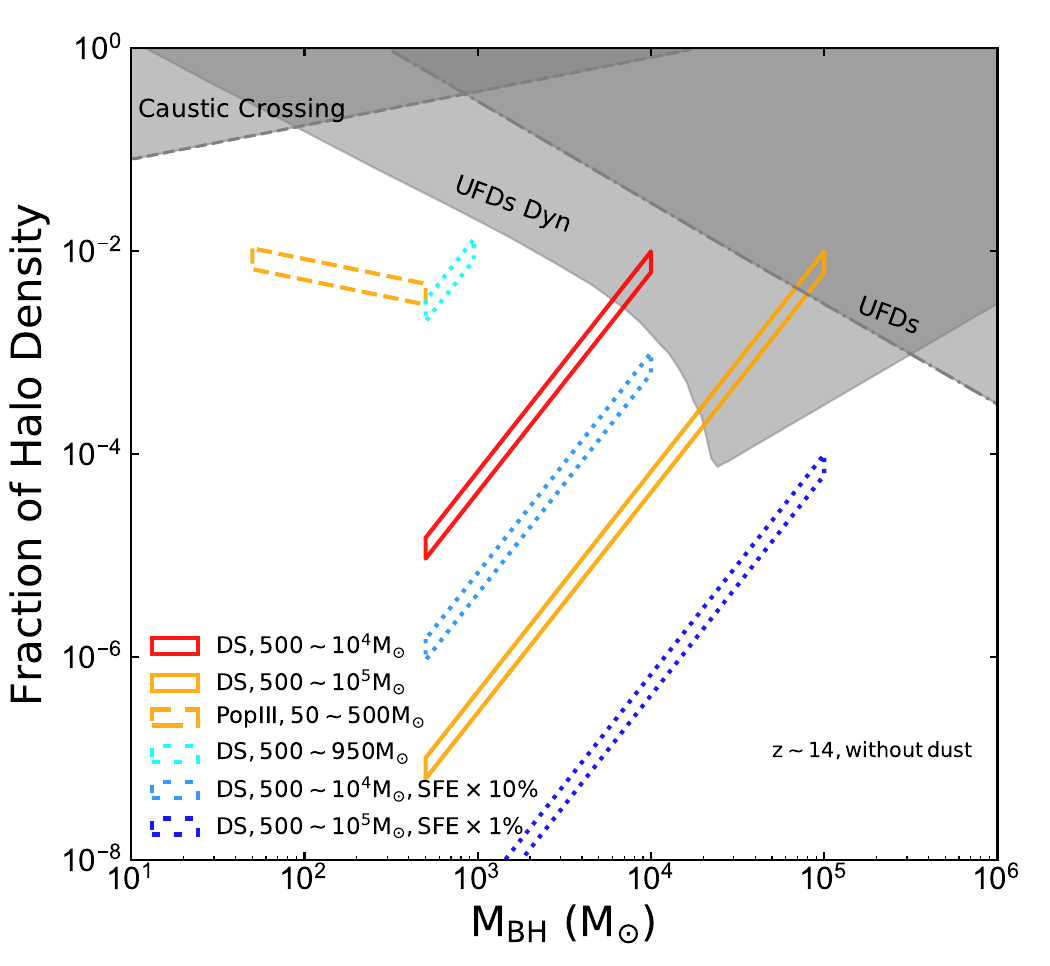}
\caption{\label{fig:Fhalo} 
Constraints on the halo mass fraction of dark stars. We adopt the results derived from various sources, including caustic crossings of strong lensing arcs \citep{2018PhRvD..97b3518O}, ultra-faint dwarf galaxies (UFDs) \citep{2016ApJ...824L..31B} and dynamic heating of UFDs \citep{2023arXiv231107654G}. The colourful boxes in solid represent the fraction results of dark stars' relic BHs of the best-fit SFE of JWST UV LFs $z\sim14$ data, corresponding to the deep red band in Figure \ref{fig:SFE_withDS} of Appendix \ref{app:SFEpost}. The orange dashed box is the mass fraction of the Pop \uppercase\expandafter{\romannumeral3} stars model within the mass range of $50\sim 500\, \rm M_{\odot}$. The dotted boxes are the allowed mass fraction of dark stars in the lower mass range ($500\sim 945\, \rm M_{\odot}$) or SFE ($\times 10\%$ within $500\sim 10^4\, \rm M_{\odot}$ and $\times 1\%$ within $500\sim 10^5\, \rm M_{\odot}$). } 
\end{figure}

After the death of those dark stars, the black holes formed and can be detected as a source of MACHOs potentially.
Those constraints derived from various experiments \citep{2016ApJ...824L..31B,2018PhRvD..97b3518O,2023arXiv231107654G} are depicted in Figure.~\ref{fig:Fhalo}.

The transformation from SFE and IMF to the halo fraction of black holes is estimated with the best-fitted effective SFEs of dark stars. The best-fitted effective SFEs of dark stars ($\epsilon_{\rm DS,eff}$) correspond to the max likelihood curves in the effective halo mass regions in Figure \ref{fig:SFE_withDS} (red or blue in case of with or without dust). 
The mass fraction of the relic BHs can be estimated via (i.e., following \cite{Carr:2020gox})
\begin{equation}\label{eq:F_halo_BH}
	\psi(m) = \epsilon_{\rm DS,eff} f_{\rm b} \left(\frac{\xi_{0} \phi(m)m^{2}}{M_{\rm tot}} \right),
\end{equation}
where $M_{\rm tot}$ is the total mass of dark stars in Eq. (\ref{eq:M_tot}).

The solid boxes in Figure~\ref{fig:Fhalo} are different halo fractions of best-fitted SFEs of different mass ranges of IMFs ($500\sim 10^4\rm M_{\odot}$ and $500\sim 10^5\rm M_{\odot}$) in z$\sim$14 UV LF dataset in case of without dust, corresponding to the deep red band in Figure \ref{fig:SFE_withDS}. The orange dashed box is the mass fraction of the Pop \uppercase\expandafter{\romannumeral3} within the mass range of $50\sim 500\, \rm M_{\odot}$ that can explain the SFE excesses at $z>10$. The dotted boxes are the allowed mass fraction of dark stars in the lower range of mass or SFE according to the constraints of MACHOs. 
The halo fraction of black holes from dark stars in conditions with or without dust is similar, thus we show only z$\sim$14 result in the case of without dust.

It is worth noting that the viability of a substantial fraction of very massive dark stars (i.e., $\geq 10^{4}M_\odot$) may face challenges imposed by some constraints. As shown in the red and orange solid boxes in Figure~\ref{fig:Fhalo}, the relic black holes of dark stars with the top-heavy IMF have been excluded by the constraints of MACHOs. Due to the top-heavy IMF, the population of the dark stars is dominated by supermassive members up to $10^4$ or $10^5\rm M_{\odot}$. Therefore, excluding the fraction of the massive part means the whole population is excluded by the constraints. To avoid the current constraints, the mass range of dark stars needs to be limited to $500\sim 945\, \rm M_{\odot}$ (dotted cyan box in Figure \ref{fig:Fhalo}). The dotted sky-blue (deep-blue) box in the mass range $500\sim 10^4\, \rm M_{\odot}$ ($500\sim 10^5\, \rm M_{\odot}$) is the halo fraction of dark stars $1/10$ ($1/100$) lower than the halo fraction of the best-fitted SFE result. In that case, the SFE of the dark star needs to be much lower than the required values that can explain the JWST observations.
However, the dark star mass is higher than $\sim 10^3 \rm \, M_{\odot}$ from fitting the high-redshift-galaxy spectrum when the dark stars are considered as the primary UV radiation sources (see Figure~\ref{fig:DMmass_DSmass}), the detail is described in Appendix \ref{app:DS}. Hence, the excess of SFE can not be explained by the dark stars, because there is no appropriate mass range allowed in MACHO constraints and spectrum fitting.

Although the estimated mass ranges of dark stars derived from UV LFs have been nearly excluded, dark stars still have a slim chance of survival. When the formation efficiency of dark stars becomes lower, such constraints from MACHO become less effective. For example, in a $10^{10}\, \rm M_{\odot}$ dark matter halo, the mass of the host galaxy is $10^8\,\rm M_{\odot}$ and the fraction of halo density of a dark star with mass $\sim 10^5 \, \rm M_{\odot}$ is $\sim 10^{-5}$, 
which is much lower than the current MACHO constraints. Therefore, the dark star can form in high-redshift galaxies indeed, but it is hard to explain the excess of SFE in JWST UV LF data.

In Section~\ref{sect:SFE}, we assumed that Pop \uppercase\expandafter{\romannumeral3} stars and dark stars have the same SFE model but different IMFs.  
The Pop \uppercase\expandafter{\romannumeral3} stars are expected to form in the dense region at high redshift.
It may be possible to explain the excess of SFE without extra limitations from MACHOs.
The solid boxes in Figure \ref{fig:Fhalo} illustrate the best-fitted fractions of Pop \uppercase\expandafter{\romannumeral3} stars. However, given the current lack of clarity regarding the IMF and other specifics of Pop \uppercase\expandafter{\romannumeral3} star formation, this analysis serves as a preliminary exploration of the data. For a more robust conclusion, future data and simulations are needed. 

\section{Discussion}
\label{discussion}

In this study, we have evaluated the potential contribution of massive dark stars to the JWST UV LFs at extremely high redshifts. Our findings indicate that massive dark stars, with masses exceeding $1000 \, \rm M_{\odot}$, which capture Weakly Interacting Massive Particles (WIMPs) as dark matter, could serve as a UV source in galaxies. This population could potentially account for the excess in star formation efficiency (SFE) observed at redshifts $z \sim 11-14$.
Nevertheless, current astrophysical constraints on the fraction of black holes within dark matter halos, derived from strong lensing perturbations and galactic dynamics, impose stringent bounds on the formation efficiency of dark stars. Consequently, we propose that dark stars likely contribute a minor fraction to the observed SFE excesses. 

Note that some interesting models have been proposed to account for the observed excesses in SFE. These include the role of early dark energy \citep{2024EPJP..139..711W,2023JCAP...10..072A} and dynamical dark energy \cite{Menci:2024rbq}, which might have influenced the expansion rate of the universe and thus the formation of structures. Primordial black holes \citep{2024SCPMA..6709512Y,2023arXiv230605364S} are another candidate that could contribute to the gravitational potential wells necessary for star formation.
Furthermore, the nature of dark matter itself is a subject of ongoing debate \citep{2023Astro...2...90N}. Warm dark matter \citep{2024RAA....24a5009L} and ultra-light dark matter \citep{2023ApJ...947...28G,2024PhLB..85839062B} are alternatives to the cold dark matter paradigm, potentially affecting the distribution and dynamics of matter in the early universe. Cosmic strings \citep{2023PhRvD.108d3510J,2023SCPMA..6620403W}, relics from phase transitions in the early universe, could also play a role in structure formation.
In addition to these new physics scenarios, other conventional astrophysical mechanisms should not be overlooked, such as feedback-free processes \citep{2023MNRAS.523.3201D,2024A&A...690A.108L} and Population III star formation \citep{2024MNRAS.527.5929Y,2022ApJ...938L..10I}. 
Distinguishing between these possibilities will require further theoretical modeling and more precise observational data.

\cite{Iocco:2024rez} demonstrated that the dark stars with masses as large as $10^6\, \rm M_{\odot}$ at $z\sim 13$ could in principle generate significant UV luminosity. However, they concluded that such objects are unlikely to explain the brightness JWST UV LFs due to the need for extremely high dark matter density and accretion rate. In this work, we have used the observed spectrum of the $z\sim 13$ galaxy JADES-GS-z13-0, combined with the Hertzsprung-Russell characteristics of dark stars, to constrain the optimal temperature and luminosity contribution of a hypothetical dark star population. Our fitting analysis favors higher dark star surface temperatures, implying either more massive dark stars or heavier WIMPs.
Nevertheless, when compared against existing MACHO constraints, we find that the number of black holes required to explain the JWST UV LFs via dark stars would exceed observational limits. This strongly disfavors dark stars as the dominant origin of the UV luminosity observed at the highest redshifts. 


\section{Acknowledgments}
We thank the reviewers for their helpful comments. We also thank Fabio Iocco,  Steven Finkelstein, Qiao Li, Yong-Jia Huang, Tian-Peng Tang, Yue-Lin Sming Tsai, Luca Visinelli, Hao-Jing Yan, Hai-Bo Yu, Qiang Yuan, Chi Zhang, Tian-Ci Zheng and Hao Zhou for their helpful discussions. This work is supported by the National Key Research and Development Program of China (No.2022YFF0503304), the Natural Science Foundation of China (No.11921003), the China Postdoctoral Science Foundation (No.2023TQ0355), and the New Cornerstone Science Foundation through the XPLORER PRIZE. G.-W. Yuan also acknowledges support from the University of Trento and
the Provincia Autonoma di Trento through the UniTrento Internal Call
for Research 2023 grant “Searching for Dark Energy off the beaten track” (DARKTRACK, grant agreement no.E63C22000500003, PI: Sunny Vagnozzi).

%



\software{astropy \citep{2013A&A...558A..33A,2018AJ....156..123A,2022ApJ...935..167A}, CAMB \citep{2000ApJ...538..473L}, pymultinest \citep{2014A&A...564A.125B}, matplotlib \citep{Hunter:2007}, numpy \citep{harris2020array}, scipy \citep{2020SciPy-NMeth}, bagpipes \citep{2018MNRAS.480.4379C, 2019MNRAS.490..417C}.}



\appendix


\label{app}

\section{UV Luminosity-to-SFR Conversion Factor ${\mathcal K}_{\rm UV}$}\label{app:UVLF}

The ${\mathcal K}_{\rm UV}$ can be calculated by two methods: spectrum integrations or mass-to-luminosity ratio ($M/L$). Previously, \cite{2023ApJS..265....5H} used spectral synthesis code to search for Pop \uppercase\expandafter{\romannumeral3} stars in the first galaxies. The $M/L$ is also used to roughly estimate the luminosity arising from the IMFs in galaxies \citep{2024MNRAS.527.5102V}.

To compare the dark stars and Pop \uppercase\expandafter{\romannumeral3} stars at high redshifts, we calculate the $M/L$ of dark stars and Pop \uppercase\expandafter{\romannumeral3} stars to derive the UV Luminosity-to-SFR conversion factor.
\begin{equation}\label{MLR}
	{\rm M/L}=\frac{M_{\rm tot}}{L_{\rm tot}},
\end{equation}

The total mass of a given IMF is
\begin{equation}\label{eq:M_tot}
	M_{\rm tot}= \int^{m_{\rm up}}_{m_{\rm low}} \xi_{0}  m\phi(m) \, dm,
\end{equation}
where $m_{\rm up}$ is the upper truncation of stellar mass, $m_{\rm low}$ is lower truncation of stellar mass, $\xi_{0}$ is a consistence, $\phi (m)$ is IMF. The Salpeter IMF and top-heavy IMF are in the same type $\phi(m) \propto m^{-\alpha}$ with different slope $\alpha$. The index is $\alpha=2.35$ for Salpeter IMF and is $\alpha=-0.17$ for a top-heavy IMF used in \cite{2024MNRAS.527.5102V}. In this work, we use a top-heavy IMF with $\alpha =-0.17$ for calculating $M/L$ of the dark stars. The IMF slope $\alpha =2.35$ is used for Pop \uppercase\expandafter{\romannumeral3} stars, which correspends to a conversion factor ${\mathcal K}_{\rm UV}= 2.80\times10^{-29} \rm \, \rm M_{\odot}\, yr^{-1} /(erg\, s^{-1}\, Hz^{-1})$ used in \cite{2022ApJ...938L..10I} and \cite{2023ApJ...954L..48W}. 

For the Pop \uppercase\expandafter{\romannumeral3} stars and dark stars with $m\geq 50\, \rm M_{\odot}$, the total luminosity of the population with a given IMF is
\begin{equation}\label{L_tot}
	L_{\rm tot}= \int^{m_{\rm up}}_{m_{\rm low}} \xi_{0}  L(m) \phi (m)\, dm,
\end{equation}
where $L(m)$ is the standard mass-to-luminosity relation of Pop \uppercase\expandafter{\romannumeral3} stars or dark stars. 

For dark stars and Pop \uppercase\expandafter{\romannumeral3} stars, $L(m)$ is derived from the interpolation of H-R diagram in \cite{2010ApJ...716.1397F} and \cite{2023ARA&A..61...65K}, respectively.

With the above Equations, we can calculate ${\mathcal K}_{\rm UV}$ of the dark stars:

\begin{equation}\label{eq:KUV_DS}
	\frac{{\mathcal K}_{\rm UV, DS}}{{\mathcal K}_{\rm UV, Pop\, \uppercase\expandafter{\romannumeral3}}} = \frac{(M/L)_{\rm DS}}{(M/L)_{\rm Pop\, \uppercase\expandafter{\romannumeral3}}},
\end{equation}

Table.~\ref{tab:KUV} shows the UV luminosity-to-SFR conversion factor ${\mathcal K}_{\rm UV}$ of stars, Pop \uppercase\expandafter{\romannumeral3} stars and dark stars. The values of Pop \uppercase\expandafter{\romannumeral2}/\uppercase\expandafter{\romannumeral1} stars and Pop \uppercase\expandafter{\romannumeral3} stars are taken from \cite{2022ApJ...938L..10I}. 
For dark stars, we calculated the values of two IMFs with the above method. Following \cite{2022ApJ...938L..10I}, we also gave the values of $\eta_{\rm UV}$ and $\epsilon_{\rm \star,rad}$ in the Table.~\ref{tab:KUV}.
\cite{2022ApJ...938L..10I} defined the $\eta_{\rm UV}$ as $\eta_{\rm UV}= \frac{L_{{\rm UV},\nu_{0}}}{SFR}$, where $\nu_{0}\simeq 8.3\, \rm eV$ corresponds to the characteristic UV wavelength $\lambda_{0}=1500 \rm\, \AA$. The $\epsilon_{\rm \star,rad}=\frac{L_{\rm UV}}{{\rm SFR}\cdot c^2}$ is defined as the UV radiative efficiency of star formation for a given
SFR by \cite{2022ApJ...938L..10I}.

\begin{table*}[htbp]
\caption{\label{tab:KUV}
UV luminosity-to-SFR conversion factor of stars, Pop \uppercase\expandafter{\romannumeral3} stars and dark stars.
}
\begin{ruledtabular}
\begin{tabular}{l cccccccc}
\textrm{Population Type} &
\textrm{IMF Type} &
\textrm{$m_{\rm low}$} &
\textrm{$m_{\rm up}$} &
\textrm{$\alpha$} & 
\textrm{$Z$} &
\textrm{${\mathcal K}_{\rm UV}$}&
\textrm{$\eta_{\rm UV}$} \textsuperscript{a} &
\textrm{$\epsilon_{\rm \star,rad}$ \textsuperscript{b}} \\
\textrm{ } &
\textrm{ } &
\textrm{$\rm M_{\odot}$} &
\textrm{$\rm M_{\odot}$} &
\textrm{ } & 
\textrm{$\rm Z_{\odot}$} &
\textrm{$\rm \frac{M_{\odot}\, yr^{-1}}{(erg \, s^{-1} Hz^{-1})}$}&
\textrm{$\rm \frac{erg \, s^{-1} Hz^{-1}}{(M_{\odot}\, yr^{-1})}$}&
\textrm{ } \\
\colrule
Stars \textsuperscript{c}
& Salpeter & $1\times 10^{-1}$ & $1\times 10^{2}$ & 2.35 & 0.02 & $1.26\times 10^{-28}$ & $7.94\times10^{27}$ & $2.79\times10^{-4}$ \\
Stars \textsuperscript{c} & Salpeter & $1\times 10^{-1}$ & $1\times 10^{2}$ & 2.35 & 0.0004 & $1.07\times10^{-28} $ & $9.32\times10^{27}$ & $3.28\times10^{-4}$ \\
Pop \uppercase\expandafter{\romannumeral3} \textsuperscript{c} & Salpeter & $5\times 10^{1}$ & $5\times 10^{2}$ & 2.35 & 0 &  $2.80\times10^{-29}$ & $3.57\times10^{28}$ & $1.26\times10^{-3}$ \\  
\hline
DS, w Cap  $m_{\chi}=10$ Gev& power-law & $5\times 10^{2}$ & $1\times 10^5$ & -0.17 & 0 & $2.79\times10^{-29} $ & $ 3.59\times10^{28}$ & $ 1.27\times10^{-3}$ \\
DS, w Cap  $m_{\chi}=100$ Gev & power-law & $5\times 10^{2}$ & $1\times 10^5$ & -0.17 & 0 & $ 2.59\times10^{-29}$ & $ 3.86\times10^{28}$ & $ 1.36\times10^{-3}$ \\
DS, w Cap  $m_{\chi}=1$ Tev & power-law & $5\times 10^{2}$ & $1\times 10^5$ & -0.17 & 0 & $ 2.70\times10^{-29}$ & $ 3.70\times10^{28}$ & $ 1.31\times10^{-3}$ \\
\cline{4-9}
DS, w Cap  $m_{\chi}=10$ Gev& power-law & $5\times 10^{2}$ & $1\times 10^4$ & -0.17 & 0 & $ 3.88\times10^{-29}$ & $ 2.58\times10^{28}$ & $ 9.10\times10^{-4}$ \\
DS, w Cap  $m_{\chi}=100$ Gev & power-law & $5\times 10^{2}$ & $1\times 10^4$ & -0.17 & 0 & $ 3.59\times10^{-29}$ & $ 2.78\times10^{28}$ & $ 9.82\times10^{-4}$ \\
DS, w Cap  $m_{\chi}=1$ Tev & power-law & $5\times 10^{2}$ & $1\times 10^4$ & -0.17 & 0 & $ 3.30\times10^{-29}$ & $ 3.03\times10^{28}$ & $ 1.70\times10^{-3}$ \\
\hline
DS, wo Cap  $m_{\chi}=10$ Gev& power-law & $5\times 10^{2}$ & $1\times 10^5$ & -0.17 & 0 & $ 2.28\times10^{-29}$ & $ 4.38\times10^{28}$ & $ 1.55\times10^{-3}$ \\
DS, wo Cap  $m_{\chi}=100$ Gev & power-law & $5\times 10^{2}$ & $1\times 10^5$ & -0.17 & 0 & $ 2.58\times10^{-29}$ & $ 3.88\times10^{28}$ & $ 1.37\times10^{-3}$ \\
DS, wo Cap  $m_{\chi}=1$ Tev & power-law & $5\times 10^{2}$ & $1\times 10^5$ & -0.17 & 0 & $ 2.76\times10^{-29}$ & $ 3.62\times10^{28}$ & $ 1.28\times10^{-3}$ \\
\cline{4-9}
DS, wo Cap  $m_{\chi}=10$ Gev& power-law & $5\times 10^{2}$ & $1\times 10^4$ & -0.17 & 0 & $ 1.63\times10^{-29}$ & $ 6.14\times10^{28}$ & $ 2.17\times10^{-3}$ \\
DS, wo Cap  $m_{\chi}=100$ Gev & power-law & $5\times 10^{2}$ & $1\times 10^4$ & -0.17 & 0 & $ 2.87\times10^{-29}$ & $ 3.48\times10^{28}$ & $ 1.23\times10^{-3}$ \\
DS, wo Cap  $m_{\chi}=1$ Tev & power-law & $5\times 10^{2}$ & $1\times 10^4$ & -0.17 & 0 & $ 3.12\times10^{-29}$ & $ 3.21\times10^{28}$ & $ 1.13\times10^{-3}$ \\
\end{tabular}
\end{ruledtabular}
\begin{tablenotes}
 \item[a] \textsuperscript{a} \cite{2022ApJ...938L..10I} defined the $\eta_{\rm UV}$ as $\eta_{\rm UV}= \frac{L_{{\rm UV},\nu_{0}}}{SFR}$, where $\nu_{0}\simeq 8.3\, \rm eV$ corresponds to the characteristic UV wavelength $\lambda_{0}=1500 \rm\, \AA$. 
 \item[b] \textsuperscript{b} The $\epsilon_{\rm \star,rad}=\frac{L_{\rm UV}}{{\rm SFR}\cdot c^2}$ is defined as the UV radiative efficiency of star formation for a given SFR by \cite{2022ApJ...938L..10I}.
 \item[c] \textsuperscript{c} The values are taken from \cite{2022ApJ...938L..10I}.

\end{tablenotes}
\end{table*}

The result of ${\mathcal K}_{\rm UV}$ shows a similar UV emission capability in different WIMP dark matter masses or upper limit of dark star masses, which is in a range of $0.7\sim 1.7$ times compared with Pop \uppercase\expandafter{\romannumeral3} stars. It shows similar properties of dark star population and Pop \uppercase\expandafter{\romannumeral3} stars in the high redshift galaxies.

The ${\mathcal K}_{\rm UV}$ values of the dark stars are similar with Pop \uppercase\expandafter{\romannumeral3} stars, which is lower than Pop \uppercase\expandafter{\romannumeral2}/\uppercase\expandafter{\romannumeral1} stars. Meanwhile, the UV emission efficiency $\epsilon_{\rm \star,rad}$ of the dark stars are similar with Pop \uppercase\expandafter{\romannumeral3} stars, which is higher than Pop \uppercase\expandafter{\romannumeral2}/\uppercase\expandafter{\romannumeral1} stars. That indicates a lower SFE in fitting the UV LFs with a ${\mathcal K}_{\rm UV}$ of the dark star or Pop \uppercase\expandafter{\romannumeral3} star compared with Pop \uppercase\expandafter{\romannumeral2}/\uppercase\expandafter{\romannumeral1} stars at lower redshfit.


\section{SFE result of UV Luminosity Function Fitting}\label{app:SFEpost}

\begin{figure}[htbp]
\centering
\includegraphics[width=0.48\linewidth]{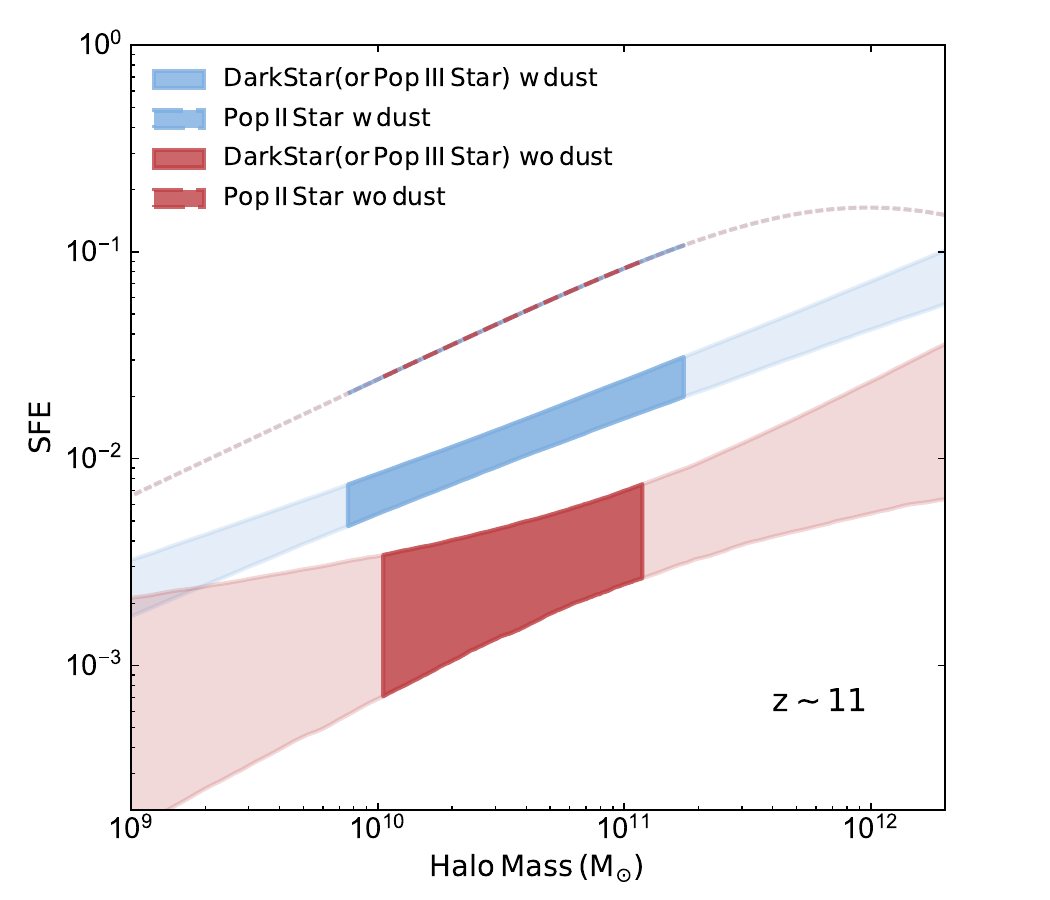}
\includegraphics[width=0.48\linewidth]{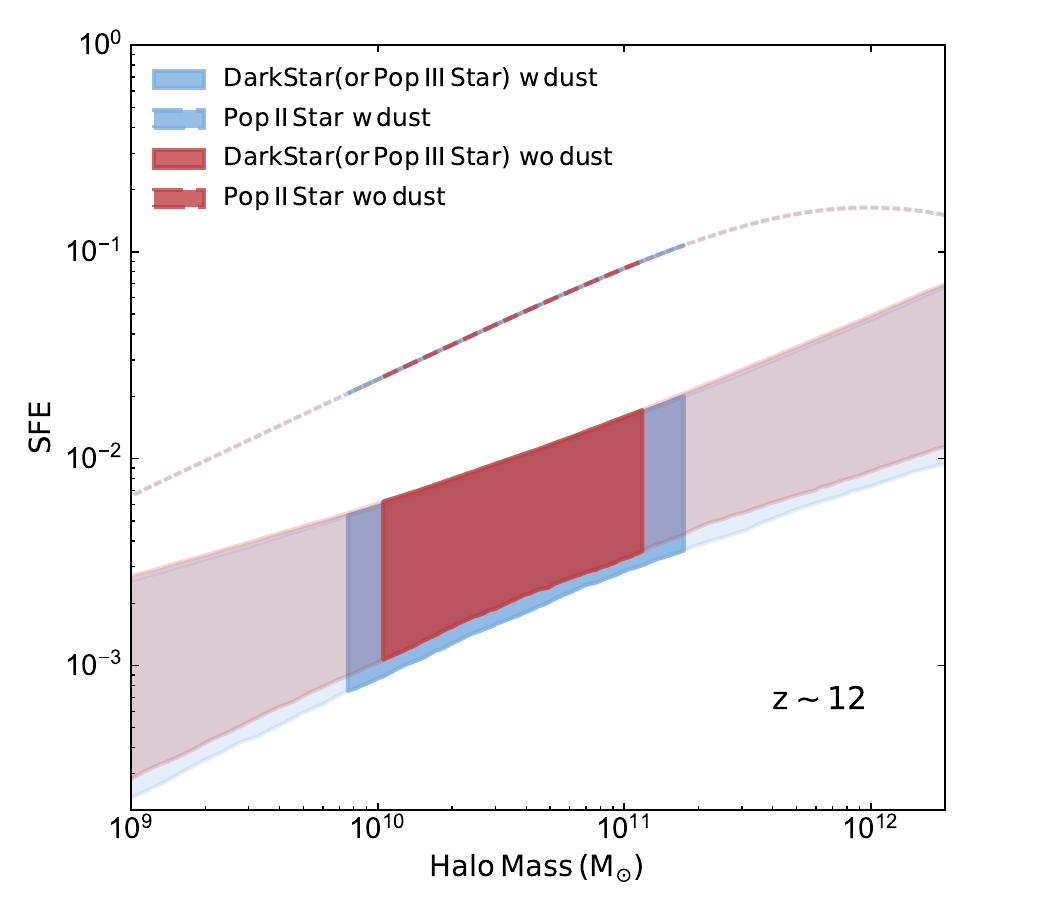}
\includegraphics[width=0.48\linewidth]{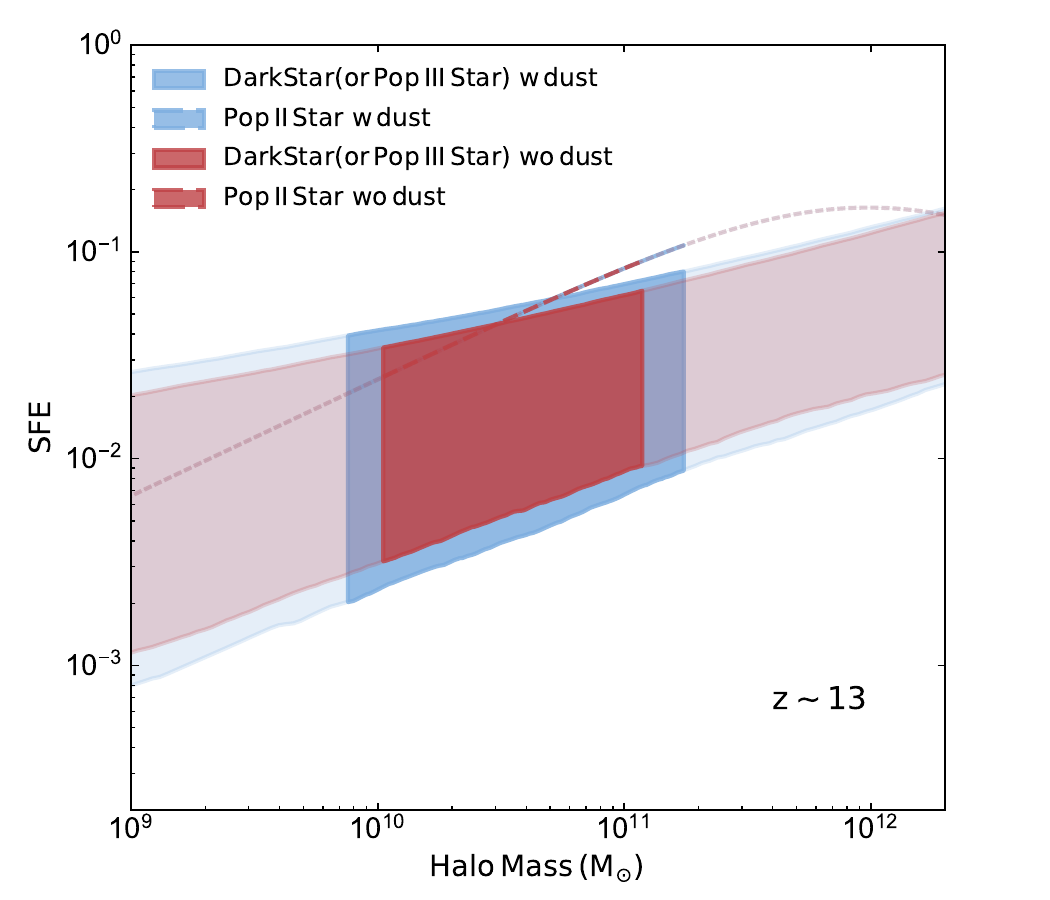}
\includegraphics[width=0.48\linewidth]{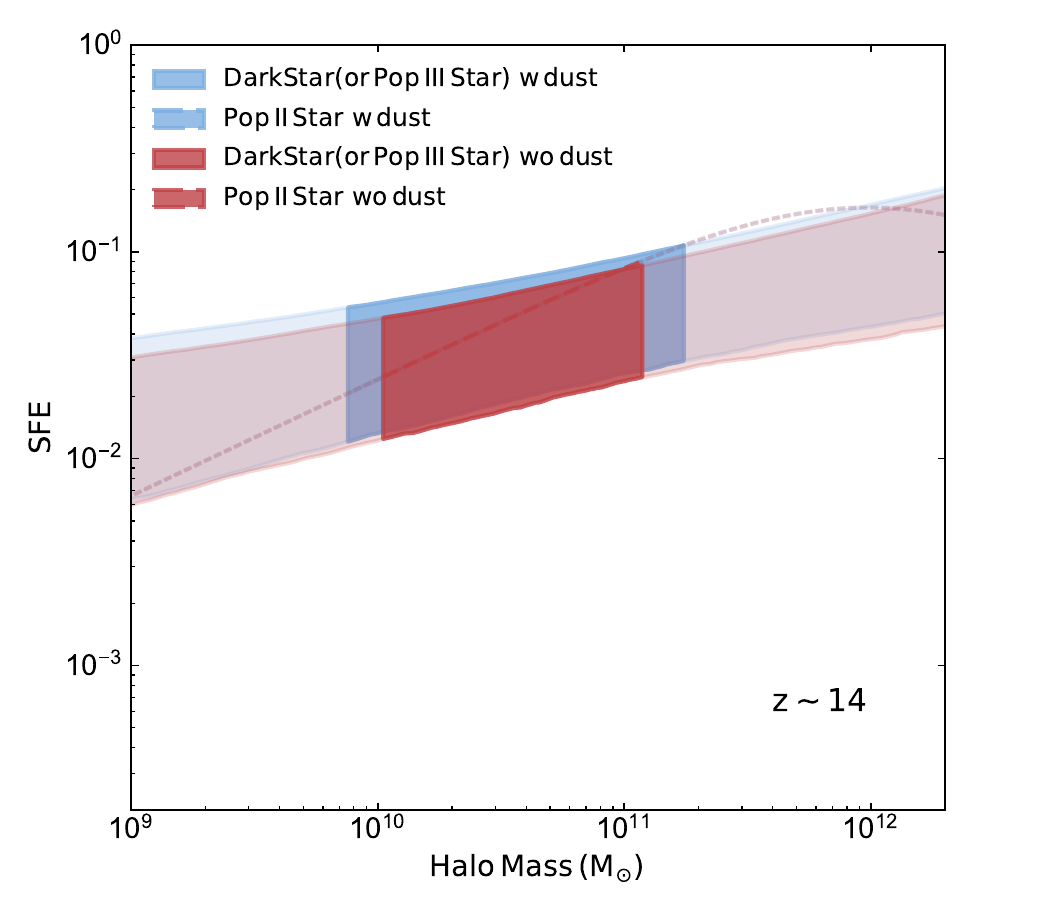}
\caption{\label{fig:SFE_withDS} Best-fit SFE models in different redshift bins with dark stars (or Pop \uppercase\expandafter{\romannumeral3} stars), with and without dust attenuation. The blue or red dashed line shows the 68\% SFE of Pop \uppercase\expandafter{\romannumeral2} stars in a range of $z\sim 4-9$ with or without dust, respectively. The blue or red region shows the 68\% best-fit SFE model of dark stars with or without dust. }
\end{figure}

At the redshift range from 4 to 9, the profiles of Pop \uppercase\expandafter{\romannumeral2} SFE are assumed to be uniform and independent with redshift increasing. As shown in Figure \ref{fig:UVLF_z4z9}, the SFE is constrained in a tight range and the UV LFs are fitted well. At higher redshift $z>10$, the contribution of the Pop \uppercase\expandafter{\romannumeral2} stars' component in total SFE follows the results in Figure \ref{fig:UVLF_z4z9}. Therefore, the dark stars (or Pop \uppercase\expandafter{\romannumeral3} stars) contribute the extra component of the total SFE.

Figure.~\ref{fig:SFE_withDS} shows best-fit SFE models in different redshift bins with dark stars (or Pop \uppercase\expandafter{\romannumeral3} stars) with dust attenuation and without dust attenuation. The blue or red dashed line shows the 68\% SFE of Pop \uppercase\expandafter{\romannumeral2} stars in a range of $z\sim 4-9$ with or without dust, respectively. The coloured region in the solid line shows the best-fit 68\% SFE model of dark stars (or Pop \uppercase\expandafter{\romannumeral3} stars). 
Our result indicates that the presence of dark stars (or Pop \uppercase\expandafter{\romannumeral3} stars) could reduce the high rate of star formation required for JWST galaxy observations, which is consistent with the current efficiency ($\sim 16\%$).

\begin{figure}[htbp]
\centering
\includegraphics[width=0.48\linewidth]{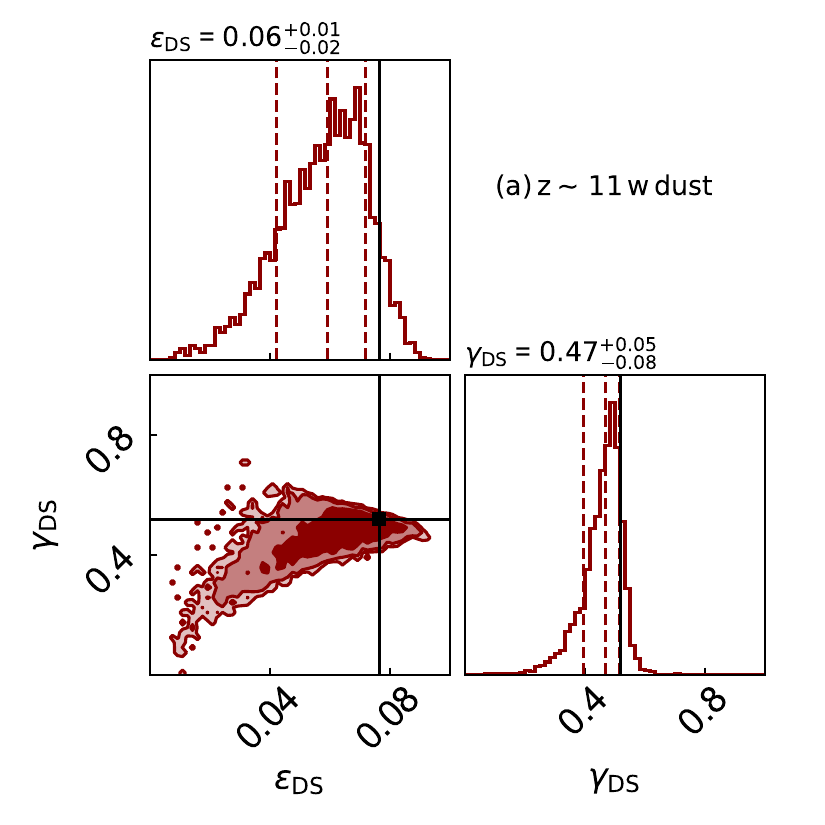}
\includegraphics[width=0.48\linewidth]{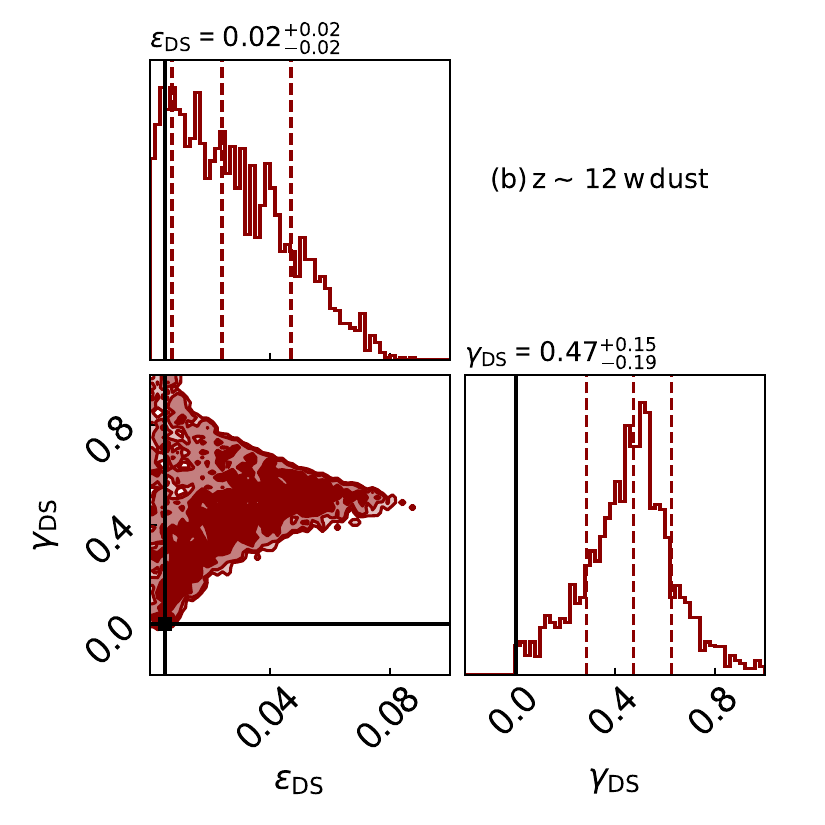}
\includegraphics[width=0.48\linewidth]{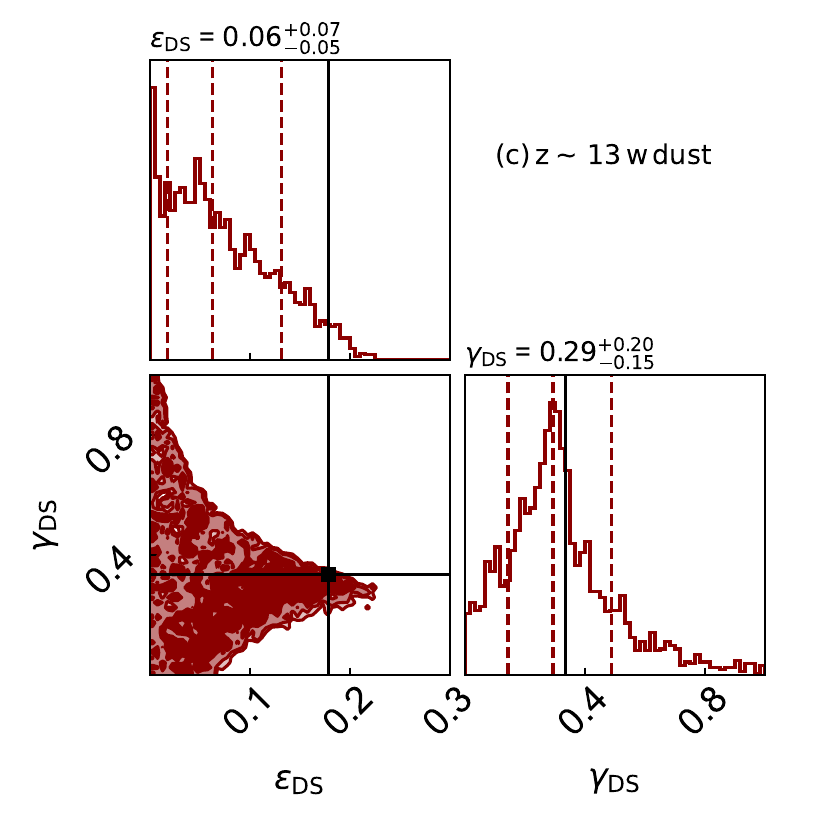}
\includegraphics[width=0.48\linewidth]{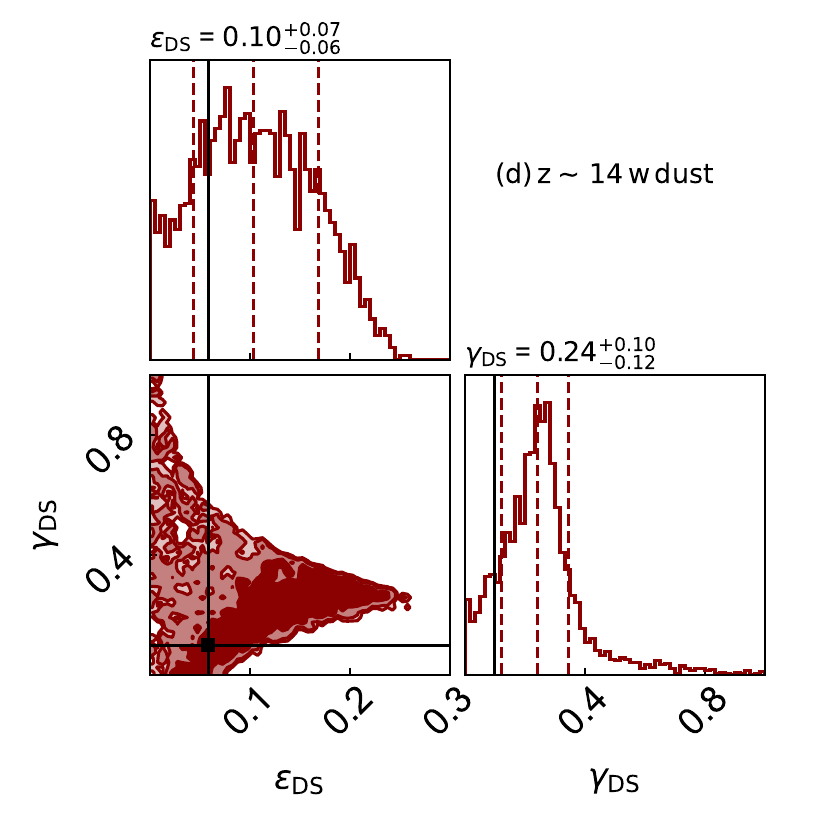}
\caption{\label{fig:posterior} 
The posteriors of dark star or Pop III star SFE model with considering dust attenuation. The darkred regions of the three depths in the corner plot are labeled with 68\%, 95\%, and 99\% confidence intervals, respectively. The black crosses are the positions of the best-fit parameters. }  
\end{figure}

\begin{figure}[htbp]
\centering
\includegraphics[width=0.48\linewidth]{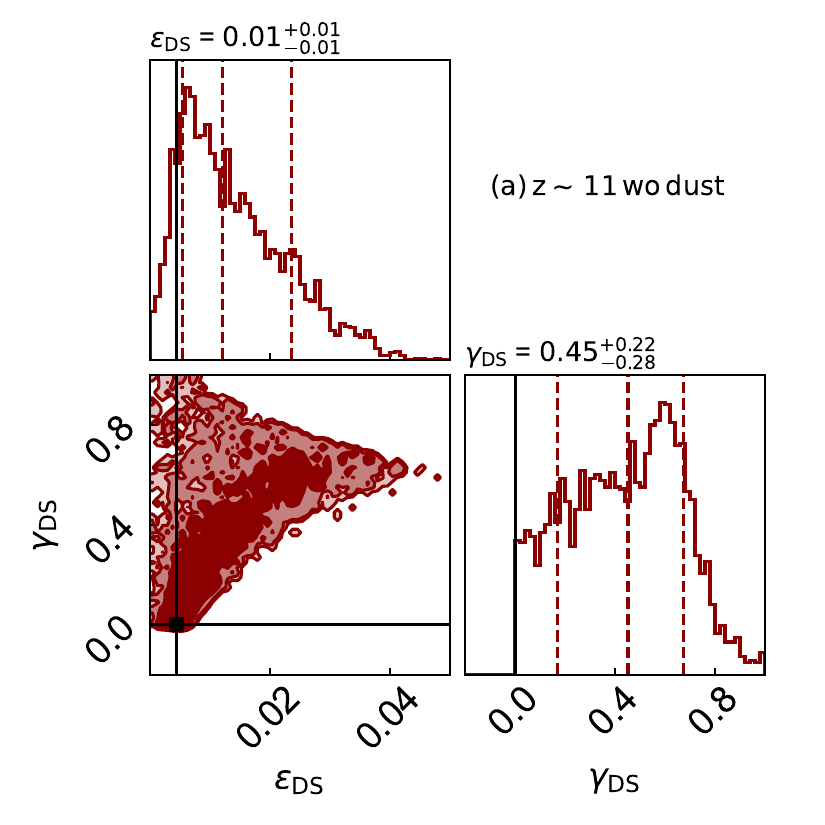}
\includegraphics[width=0.48\linewidth]{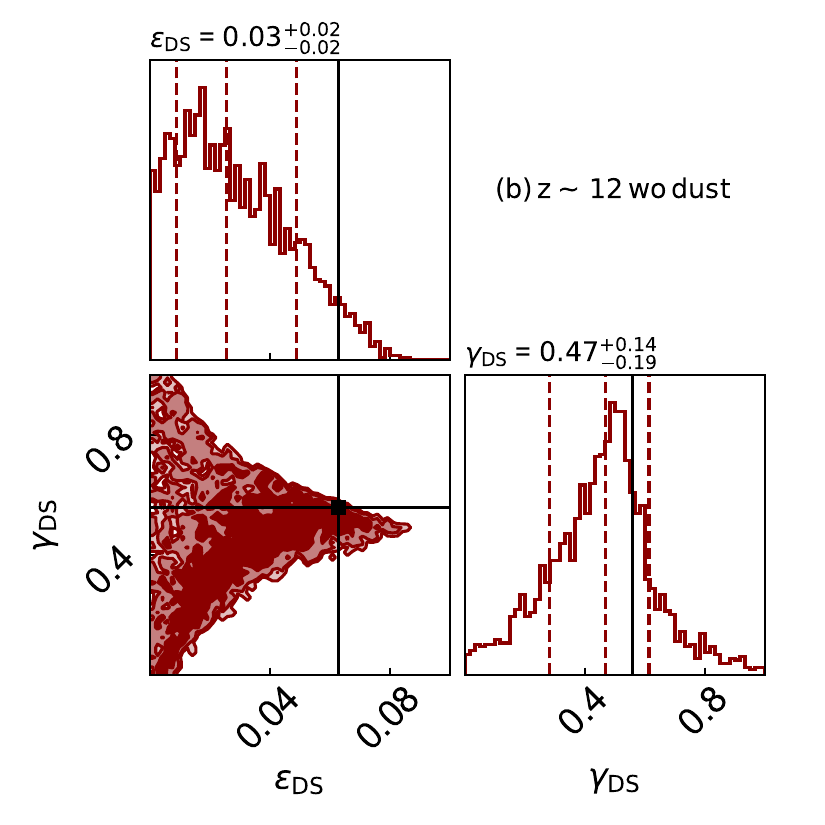}\\
\includegraphics[width=0.48\linewidth]{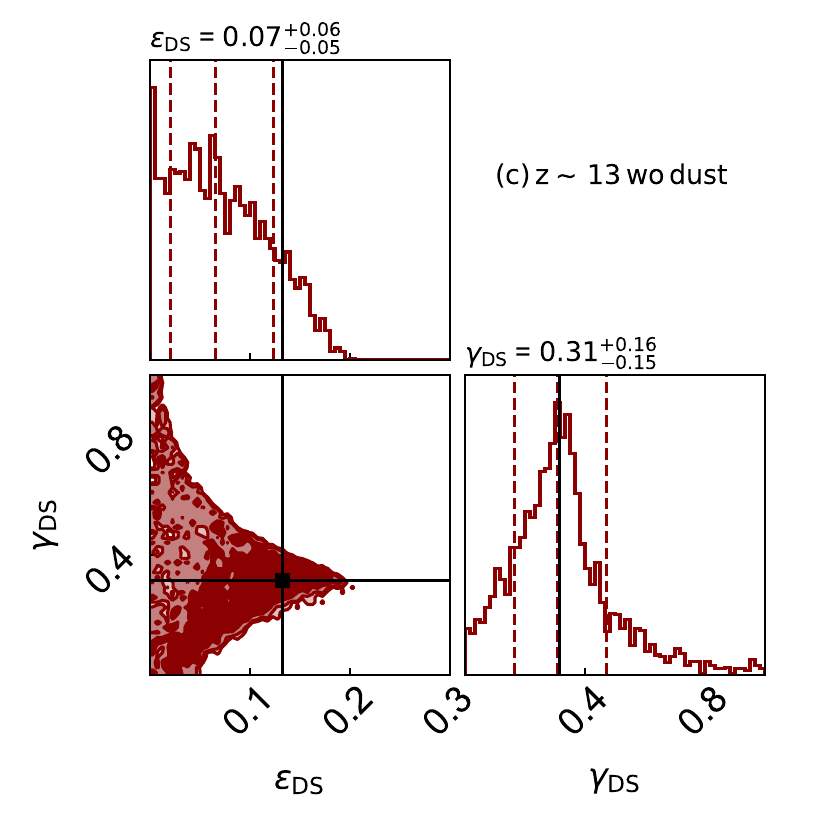}
\includegraphics[width=0.48\linewidth]{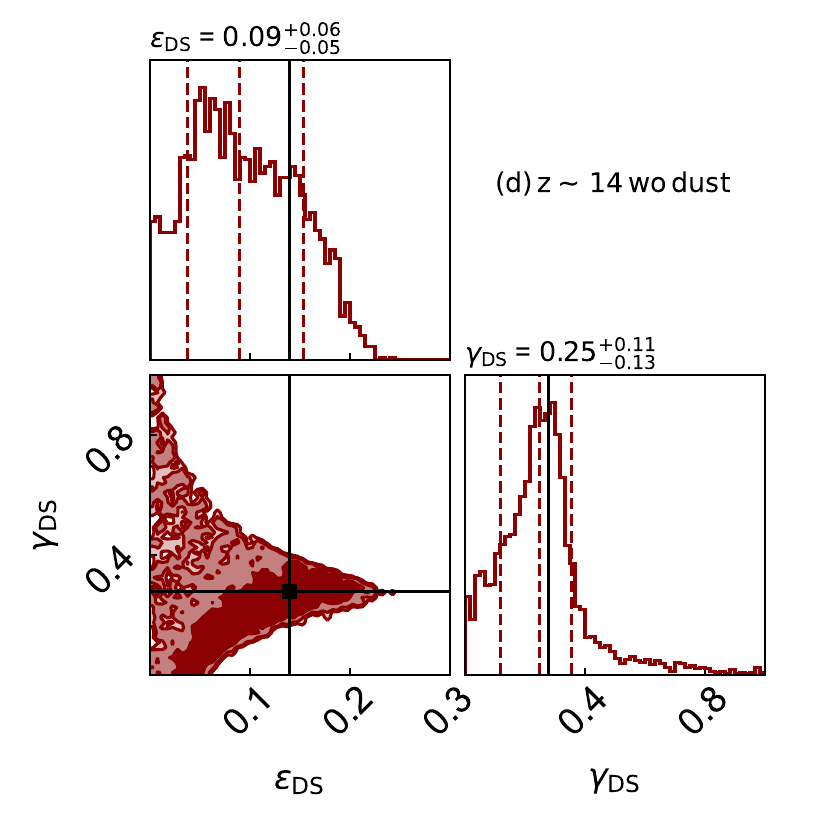}
\caption{\label{fig:posteriorwodust} 
The posteriors of dark star or Pop III star SFE model without considering dust attenuation. The darkred regions of the three depths in the corner plot are labeled with 68\%, 95\%, and 99\% confidence intervals, respectively. The black crosses are the positions of the best-fit parameters. }  
\end{figure}

We also plot the corners of the posteriors in Figures \ref{fig:posterior} and \ref{fig:posteriorwodust}. The dark red regions of the three depths in the corner plots of \ref{fig:posterior} and \ref{fig:posteriorwodust} represent 68\%, 95\%, and 99\% confidence intervals, respectively. The black crosses are the positions of the best-fit parameters.

\section{Constraint on Dark Star Properties with JWST Spectrum}
\label{app:DS}
The properties of dark stars can be effectively constrained through the fitting of spectra obtained from high-redshift galaxies at $z>12$. At such high redshift, the galaxy spectrum may be dominated by massive dark stars when the stellar populations within the galaxy consist of a mixture of traditional stars and dark stars. 
By analyzing the spectrum of high-redshift galaxies, it becomes possible to constrain both the dark matter mass and the dark stars. In this context, we perform a fitting procedure for the dark star temperature, assuming that the UV radiation is primarily governed by the presence of massive dark stars.

As a pertinent example, JADES-GS-z13-0 stands out as a spectrum-confirmed galaxy positioned at a redshift of $z=13.2$, characterized as a metal-poor young galaxy \citep{2023NatAs...7..622C}. Its NIRSpec data has been made publicly available in \cite{https://doi.org/10.17909/8tdj-8n28,2024A&A...690A.288B}, and the data reduction processes can be found in \citep{2023arXiv230602465E,2024arXiv240406531D}.  We leverage this galaxy spectrum to fit the parameters associated with dark stars and WIMPs. This dataset provides a valuable opportunity to refine our understanding of the physical characteristics of dark stars and their interply with WIMPs in the high-redshift galaxy environment.

\begin{figure}[htbp]
\centering
\includegraphics[width=0.99\linewidth]{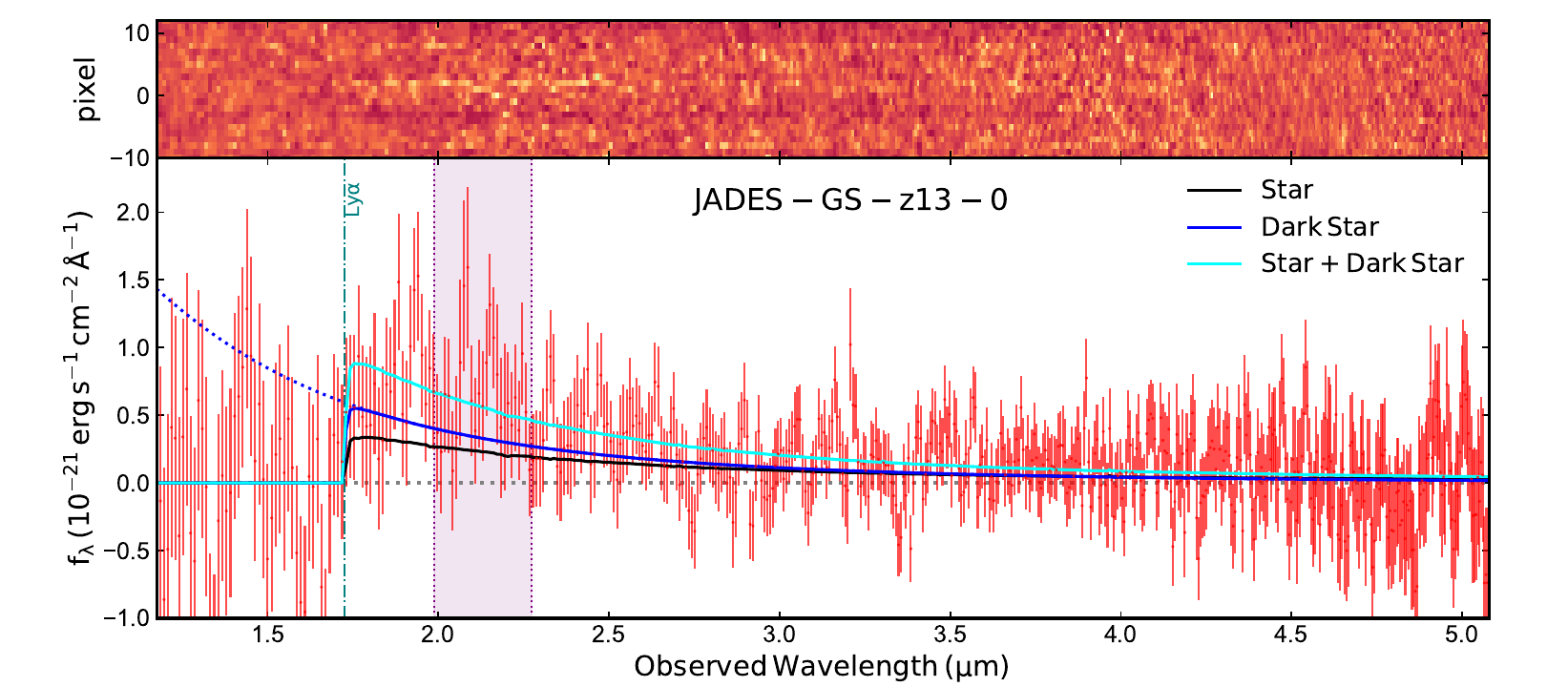}
\caption{\label{fig:JADESz13} The JWST/NIRSpec spectrum and best-fit spectrum model of JADES-GS-z13-0. The yellow solid line is young stellar component contributions in the galaxy model from the software \emph{Bagpipes} \citep{2018MNRAS.480.4379C, 2019MNRAS.490..417C}. The solid or dotted blue lines are absorbed or unabsorbed blackbody models of dark stars respectively. The cyan solid line is the best-fit total spectrum model in this work. The red error bars are the NIRSpec spectrum of the galaxy. The NIRSpec data has been released in \cite{2024A&A...690A.288B}. The purple band is a filter window at $1500 \rm \AA$ with width $200 \rm \AA$.}
\end{figure}

We rescale the dark stars' blackbody spectral model ($\rm BB_{DS}$) with varying temperatures and the young stellar spectra ($\rm YS$) from \emph{Bagpipes} to achieve a matching flux of $flux_{1500\, \rm \AA} = \rm 6.56\times 10^{21}\,\, erg\,\,s^{-1}\,\,cm^{-2}\,\,\AA^{-1}$ of the galaxy JADES-GS-z13-0 spectrum at $1500\, \rm \AA$. The flux at $1500\, \rm \AA$ is calculated using a $\sim 200\, \rm \AA$ top-hat filter, represented by the purple band in Figure~\ref{fig:JADESz13}. Subsequently, we performed a fitting procedure to determine the temperature $T$ of ${\rm BB_{DS}} (T)$ and the fraction of dark stars' UV radiation at 1500 {\r A} ($R_{\rm DS}$). The priors of $R_{\rm DS}$ and $log(T)$ are uniform distributions within the specified ranges. 

With the fitting results of the JWST UV LFs of $z>10$ galaxies, we derive the range of prior fraction for the UV radiation of dark stars at 1500 {\r A} ($R_{\rm DS}$):
\begin{equation}\label{eq:R_DS}
R_{{\rm DS},M_{\rm h}}\equiv \frac{f_{{\rm DS}, M_{\rm h}} {\mathcal K}_{\rm UV, S}} {f_{{\rm DS}, M_{\rm h}} {\mathcal K}_{\rm UV, S}+f_{{\rm S}, M_{\rm h}} {\mathcal K}_{\rm UV, DS}},
\end{equation}
where $f_{{\rm DS}, M_{\rm h}}$ is SFE of dark star at a halo mass $M_{\rm h}$ in Eq.~(\ref{eq:SFE_DS}), ${\mathcal K}_{\rm UV, S}$ is UV luminosity-to-SFR conversion factor of stars, ${\mathcal K}_{\rm UV, DS}$ is UV luminosity-to-SFR conversion factor of dark stars. For the redshift bin $z\sim13$, we set the prior of $R_{\rm DS}$ as $\sim 0-84.5\%$.

We use the galaxy spectrum Bayesian analysis tool \emph{Bagpipes} \citep{2018MNRAS.480.4379C, 2019MNRAS.490..417C}, to construct the spectrum of a galaxy with young stellar populations. For modelling the dark star spectrum, we use a blackbody model ($\rm BB_{DS}$). 
To specify the properties of the young stellar component, we assigned an age of approximately $\sim200\, \rm {Myr}$, corresponding to the onset of star formation at a redshift of $z\sim25$. The metallicity of the young stellar populations was set as ${\rm log} (Z/Z_{\odot})=-1.69$~\citep{2023NatAs...7..622C}. The nebular ionization emission is determined by the ionization parameter ${\rm log} U$, following \cite{2023NatAs...7..622C}, we estimate it with metallicity:
\begin{equation}\label{eq:metallicity-to-U}
{\rm log} U_{s} = -3.638 + 0.055 Z+ 0.68 Z^2.
\end{equation}
The total flux of galaxy at wavelength $\rm \lambda_{i}$ is given by the equation:
\begin{equation}\label{eq:fit-spec}
flux_{\rm tot,\lambda_{i}} = R_{\rm DS} flux_{{\rm BB_{DS}} (T),\lambda_{i}} + \left(1-R_{\rm DS}\right) flux_{\rm YS,{\rm \lambda_{i}}}.
\end{equation}

Although dark stars can emit UV photons with higher energy than $13.4 \rm \, eV$, the far-UV radiation will be absorbed by the neutral intergalactic medium (IGM). Thus, we calculate the accurate profile of the damping wing of the Gunn-Peterson trough
caused by a homogeneous neutral IGM. Following \cite{1998ApJ...501...15M}, the optical depth of Ly$\alpha$ damping wing absorption is described by a exponential index $\tau (\Delta\lambda)$
\begin{equation}\label{eq:G-P Trough}
\tau (\Delta\lambda) = \frac{\tau_0 R_{\alpha}}{\pi} \left(1+\delta \right)^{3/2} \int^{x_2}_{x_1} \frac{dx\, x^{9/2}}{\left(1-x \right)^2} ,
\end{equation}
where $\delta \simeq \Delta\lambda / \left[ \lambda_{\alpha}\left(1 + z_s \right) \right] $, $x_1=(1+z_n )/\left[(1+z_s)(1+\delta)\right]$, $z_s=13.20$ is the source redshift, $z_n=13.17$ is the redshift of the foreground neutral IGM that is constrained by \cite{2023NatAs...7..622C}, $x_2=(1+\delta)^{-1}$. , and the integral is given by \cite{1998ApJ...501...15M}:
\begin{equation}\label{eq:G-P Trough}
\int^{x_2}_{x_1} \frac{dx\, x^{\frac{9}{2}}}{\left(1-x \right)^2} = \frac{x^{\frac{9}{2}}}{1-x} + \frac{9}{7}x^{\frac{7}{2}} + \frac{9}{5}x^{\frac{5}{2}} +3 x^{\frac{3}{2}} +9 x^{\frac{1}{2}} -\frac{9}{2} {\rm log}\frac{1+x^{\frac{1}{2}}}{1-x^{\frac{1}{2}}} .
\end{equation}

Hence, the total flux of the galaxy spectrum model is:
\begin{equation}\label{eq:tot_spec}
	Flux_{\rm tot,\lambda_{i}} = e^{-\tau(\Delta\lambda)} flux_{\rm tot,\lambda_{i}} .
\end{equation}

In the fitting procedure, we first fitted the dark star parameters $R_{\rm DS}$ and $T$. After correcting the IGM absorption with the Eq.~(\ref{eq:tot_spec}), the total flux best fitted needs a total shift into a more well-fitted case because the above scaling of the blackbody and young stellar spectra is limited in a $\sim 200\, \rm  \AA$ top-hat filter. Then, we used a total flux amplitude to fit the best-fitted spectrum model including $BB_{\rm DS} (T)$ and young stellar. The best-fit amplitude of the total spectral model is $0.65\pm 0.04$ at $1\sigma$ confidence level.

\begin{figure}[htbp]
\centering
\includegraphics[width=0.6\linewidth]{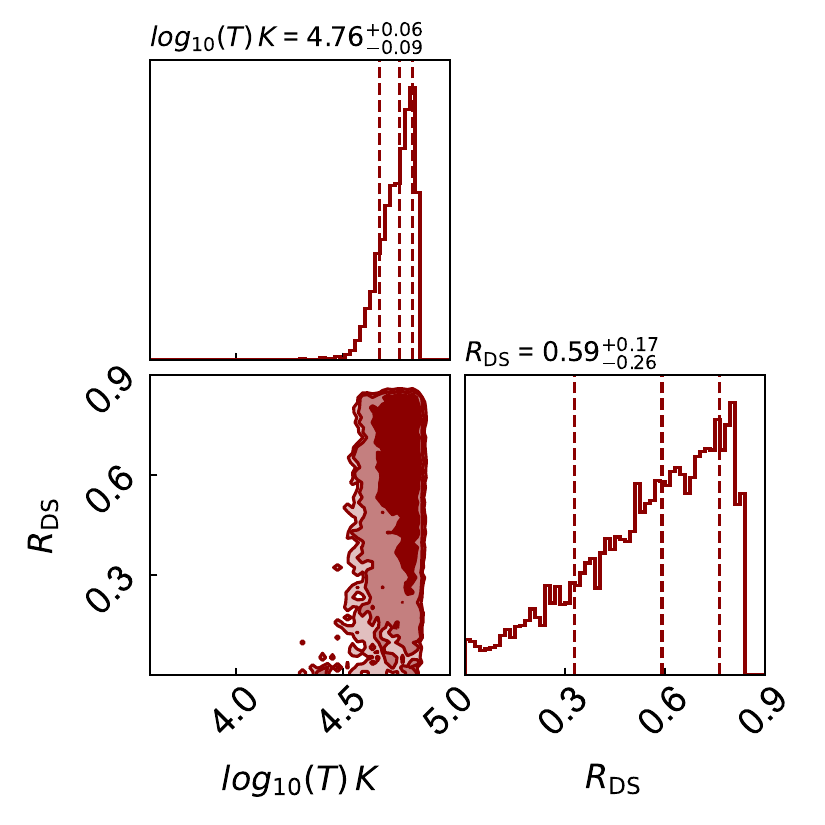}
\caption{\label{fig:JADESz13+post} The posteriors of fitting the galaxy spectrum of JADES-GS-z13-0.}
\end{figure}

\begin{figure}[htbp]
\centering
\includegraphics[width=0.6\linewidth]{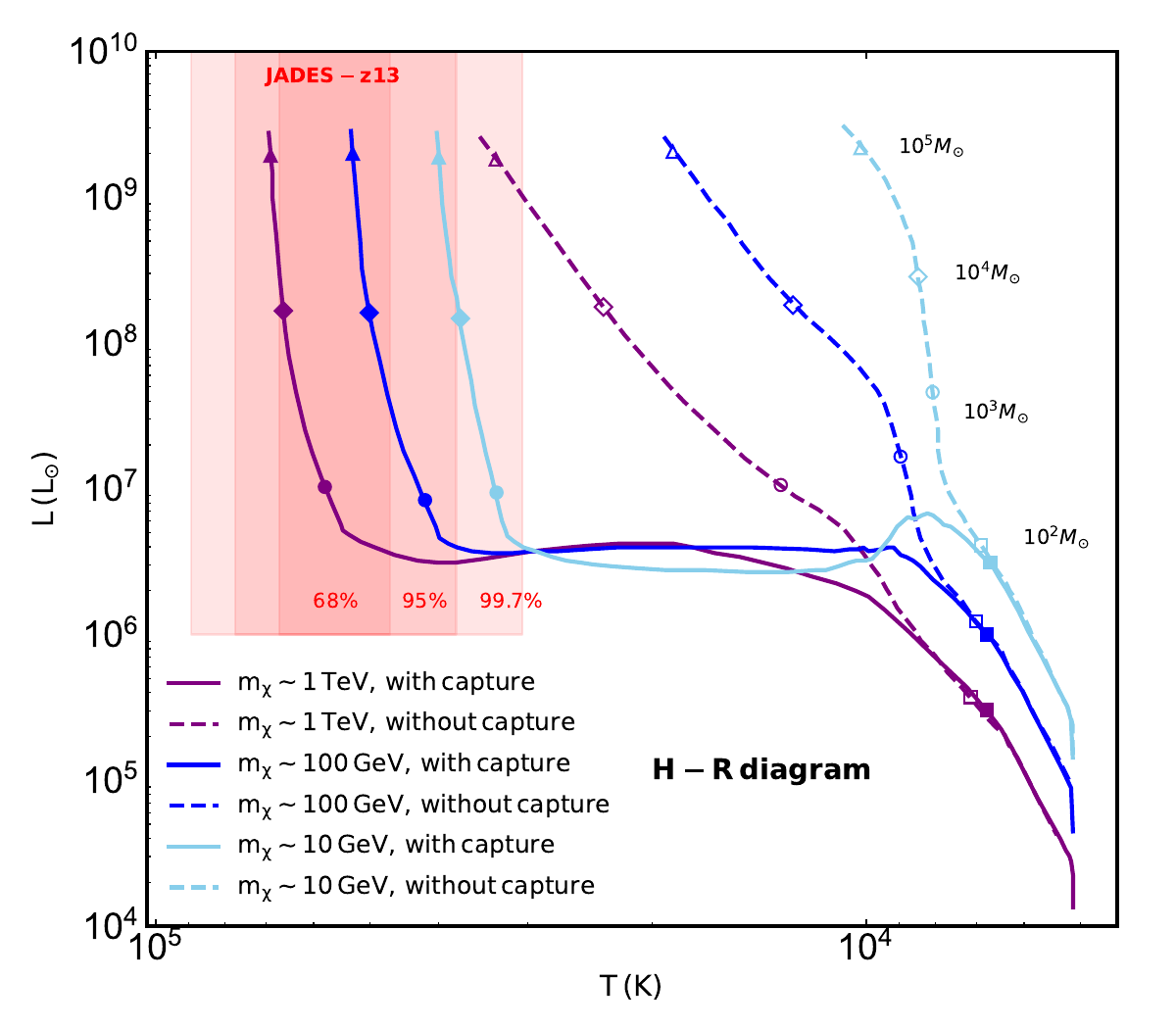}
\caption{\label{fig:JADESz13+HRdiagram} Hertzsprung-Russell (H-R) diagram of dark stars and the posteriors of dark stars' temperature with JWST spectrum of JADES-z13. The H-R diagram is reproduced from ref.~\cite{2010ApJ...716.1397F}. The red regions with different depths are 68\%, 95\% and 99.7\% posterior regions of the dark star temperature by fitting the spectrum of object JADES-z13.}
\end{figure}

In Fig~\ref{fig:JADESz13}, we present the JWST/NIRSpec spectrum and their best-fit model parameters, and the posteriors of the two parameters $R_{\rm DS}$ and $T$ are shown in Fig~\ref{fig:JADESz13+post}. Furthermore, we display the H-R diagram of dark stars and the posteriors of dark stars' temperature with the JWST spectrum of JADES-z13 in Fig~\ref{fig:JADESz13+HRdiagram}.

In Figure \ref{fig:JADESz13}, the best-fit model and the JWST/NIRSpec spectrum are presented, revealing the dominance of the dark star in the galaxy spectrum, with the stellar component contributing secondarily in the UV band. Figure \ref{fig:JADESz13+post} displays the posteriors of the two parameters $R_{\rm DS}$ and $T$, indicating a blackbody temperature of the dark stars' surface in the galaxy at $\sim 5.75 \times 10^4 \, \rm K$. Furthermore, the estimated fraction of UV radiation contributed by the dark stars in the galaxy is approximately $59\%$.

\section{The properties of WIMP Dark Matter}
\label{WIMPs}

\begin{figure}[htbp]
\centering
\includegraphics[width=0.6\linewidth]{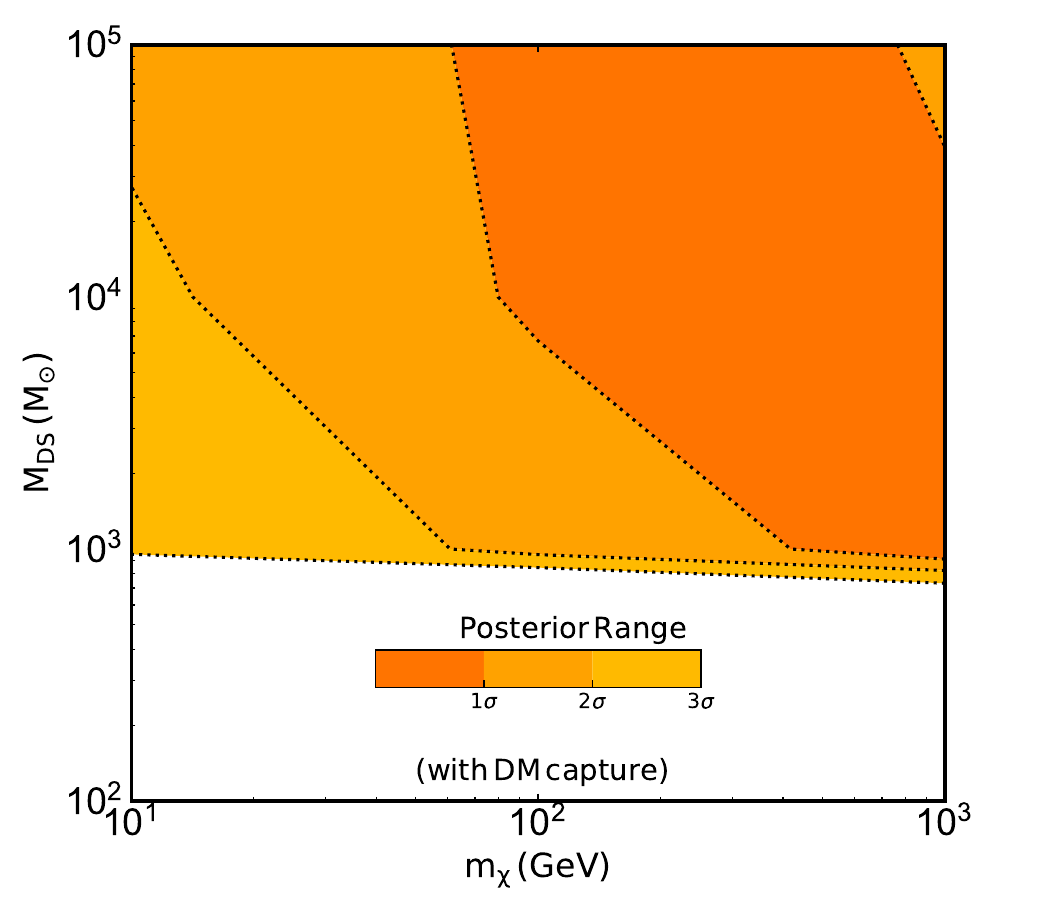}
\caption{\label{fig:DMmass_DSmass} The posterior distribution of the dark star parameter $\rm M_{DS}$ and the mass parameter $m_{\chi}$ for WIMPs. }
\end{figure}

The properties of Dark stars are intricately linked to the particle mass of WIMPs. In the study conducted by \cite{2010ApJ...716.1397F}, dark stars' physical characteristics were explored under two scenarios: one ``with capture'' and the other ``without capture''. To visualize the distribution of dark stars in the Hertzsprung-Russell (H-R) diagram, we present Figure~\ref{fig:JADESz13+HRdiagram}. 
The dark stars, with varying masses, exhibit distinct temperatures and luminosities on the H-R diagram. Leveraging this diagram, we constrain the parameters of WIMP dark matter and the mass of dark star. The results, particularly for the "with capture" scenario, are displayed in Figure~\ref{fig:DMmass_DSmass}. Notably, our finding underscore the necessity of very massive dark stars, exceeding $\sim 10^{3}M_\odot$, due to the strong suppression of the luminosity at lower masses (see also Fig.\ref{fig:JADESz13+HRdiagram} in the Appendix~\ref{app:DS}). Conversely, in the case of ``without capture", no suitable parameter region was identified.

Our results presented in Figure \ref{fig:DMmass_DSmass} indicate a preference for WIMPs with masses ranging from tens of GeV to a few TeV. Intriguingly, this range aligns with the GeV Gamma-ray excess observed in the inner Galaxy~\citep{2011PhLB..697..412H,2015PhRvD..91l3010Z}, the possible anti-proton excess~\citep{2017PhRvL.118s1101C}, and the W-boson mass anomaly~\citep{2022Sci...376..170C}. The consistent interpretation of these phenomena as the annihilation of $\sim 50-70$ GeV WIMPs~\citep{2022PhRvL.129i1802F,2022PhRvL.129w1101Z} adds support to the hypothesis that the annihilation of WIMPs can fuel the dark stars discussed in this study.


\end{CJK*}

\bibliography{sample631}{}
\bibliographystyle{aasjournal}



\end{document}